\documentclass[manuscript,screen,nonacm, authorversion, review=false, timestamp=false]{acmart} 

\AtBeginDocument{%
  }

\copyrightyear{2025}
\acmYear{2025}
\setcopyright{cc}
\setcctype{by-nc-sa}
\acmConference[CHI '25]{CHI Conference on Human Factors in Computing Systems}{April 26-May 1, 2025}{Yokohama, Japan}
\acmBooktitle{CHI Conference on Human Factors in Computing Systems (CHI '25), April 26-May 1, 2025, Yokohama, Japan}\acmDOI{10.1145/3706598.3713910}
\acmISBN{979-8-4007-1394-1/2025/04}

\usepackage{enumitem}
\usepackage{acmart-taps}

\begin{document}

\title{SocialEyes: Scaling Mobile Eye-tracking to Multi-person Social Settings}

\author{Shreshth Saxena}
\orcid{0000-0002-9237-5461}
\affiliation{
  \institution{McMaster University}
  \city{Hamilton, ON}
  \country{Canada}} 
\email{saxens17@mcmaster.ca}

\author{Areez Visram}
\affiliation{
  \institution{McMaster University}
  \city{Hamilton, ON}
  \country{Canada}}

\author{Neil Lobo}
\affiliation{
  \institution{McMaster University}
  \city{Hamilton, ON}
  \country{Canada}}

\author{Zahid Mirza}
\affiliation{
  \institution{McMaster University}
  \city{Hamilton, ON}
  \country{Canada}}

\author{Mehak Rafi Khan}
\affiliation{
  \institution{McMaster University}
  \city{Hamilton, ON}
  \country{Canada}}

\author{Biranugan Pirabaharan}
\affiliation{
  \institution{McMaster University}
  \city{Hamilton, ON}
  \country{Canada}}

\author{Alexander Nguyen}
\affiliation{
  \institution{McMaster University}
  \city{Hamilton, ON}
  \country{Canada}}

\author{Lauren K Fink}
\affiliation{
  \institution{McMaster University}
  \city{Hamilton, ON}
  \country{Canada}}
\email{finkl1@mcmaster.ca}

\renewcommand{\shortauthors}{Saxena et al.}

\begin{abstract}

  Eye movements provide a window into human behaviour, attention, and interaction dynamics. Challenges in real-world, multi-person environments have, however, restrained eye-tracking research predominantly to single-person, in-lab settings. We developed a system to stream, record, and analyse synchronised data from multiple mobile eye-tracking devices during collective viewing experiences (e.g., concerts, films, lectures). We implemented lightweight operator interfaces for real-time-monitoring, remote-troubleshooting, and gaze-projection from individual egocentric perspectives to a common coordinate space for shared gaze analysis. We tested the system in a live concert and a film screening with 30 simultaneous viewers during each of two public events (N=60).  We observe precise time-synchronisation between devices measured through recorded clock-offsets, and accurate gaze-projection in challenging dynamic scenes. Our novel analysis metrics and visualizations illustrate the potential of collective eye-tracking data for understanding collaborative behaviour and social interaction. This advancement promotes ecological validity in eye-tracking research and paves the way for innovative interactive tools.
\end{abstract}

\maketitle

\section{Introduction}

Eye contact and shared gaze are critical in social interactions and in guiding collective/joint attention. Eye gaze performs a unique dual function of perception and signalling \cite{gobel2015gazeDualFunction}, making it an inherent part of human social behaviour. Tracking when, where, and how people shift their gaze in social interactions provides insights into the dynamics of individual as well as shared attention allocation \cite{valtakari2021eye, clark1985language}, the influence of social context and co-presence \cite{foulsham2010SocialGaze}, the role of non-verbal cues in communication \cite{senju2008GazeFollowing}, and affective mechanisms such as the level of interest and engagement \cite{rahal2019GazeAffect}. Moreover, due to its unobtrusive and direct nature, there has been a long-standing interest in interactive applications of eye-tracking \cite{jacob1990CHI}. These applications range from using eye movements as a proxy for traditional computer inputs \cite{colin1987computerInput, cui2023glancewriter} to the development of innovative and creative tools \cite{creed2020creative1, hornof2014creative2, vamvakousis2016eyeharp}. Eye movements, therefore, serve as an exceptional tool for cognition research and human-computer interaction applications.

Tracking eye movements has mostly been limited to single-person settings under controlled environments due to the immobility of traditional eye-tracking hardware and the complexities of unrestricted multi-person data collection and analysis. With the recent development of affordable, mobile eye-tracking technology, it is increasingly possible to conduct eye-tracking “in the wild” (see e.g., \cite{fasold2021gaze, kulke2021implicit, reitstatter2020display, saxena2023deep, vandemoortele2018gazing, fridman2018cognitive}). For single-person use cases, mobile eye-tracking glasses are most commonly applied to collect data in unrestricted settings but a number of challenges remain unsolved for their adoption in multi-person settings. For instance, these glasses record individual gaze from the egocentric perspective of the wearer; data from all differing perspectives of wearers need to be aligned when conducting multi-person studies. To effectively study collective gaze dynamics in naturalistic social settings, it is crucial to develop automated ways to synchronise and analyse the eye-tracking data of multiple individuals.

In this paper, we look into the ubiquitous and cross-cultural social setting of collective viewing experiences where multiple people share a common gaze goal, e.g., sporting events \cite{kredel2017eye}, surgery operations \cite{tolvanen2022eye}, school classrooms \cite{jarodzka2021eye}, etc. Eye-tracking in such contexts holds the potential of studying social cognition over extended timescales and developing novel interactive tools that promote group coordination and collaboration. For instance, while we have learned much about the cognitive neuroscience of music via laboratory studies with single participants and short sound clips presented in the lab, there is a push in the field for studying musical engagement in naturalistic social contexts \cite{tervaniemi2023neuroscience}, as the types of experiences people report as deeply meaningful often involve social settings with music that evolves over long periods of time. 

\begin{figure}[htbp]
  \centering
  \includegraphics[width=0.8\linewidth]{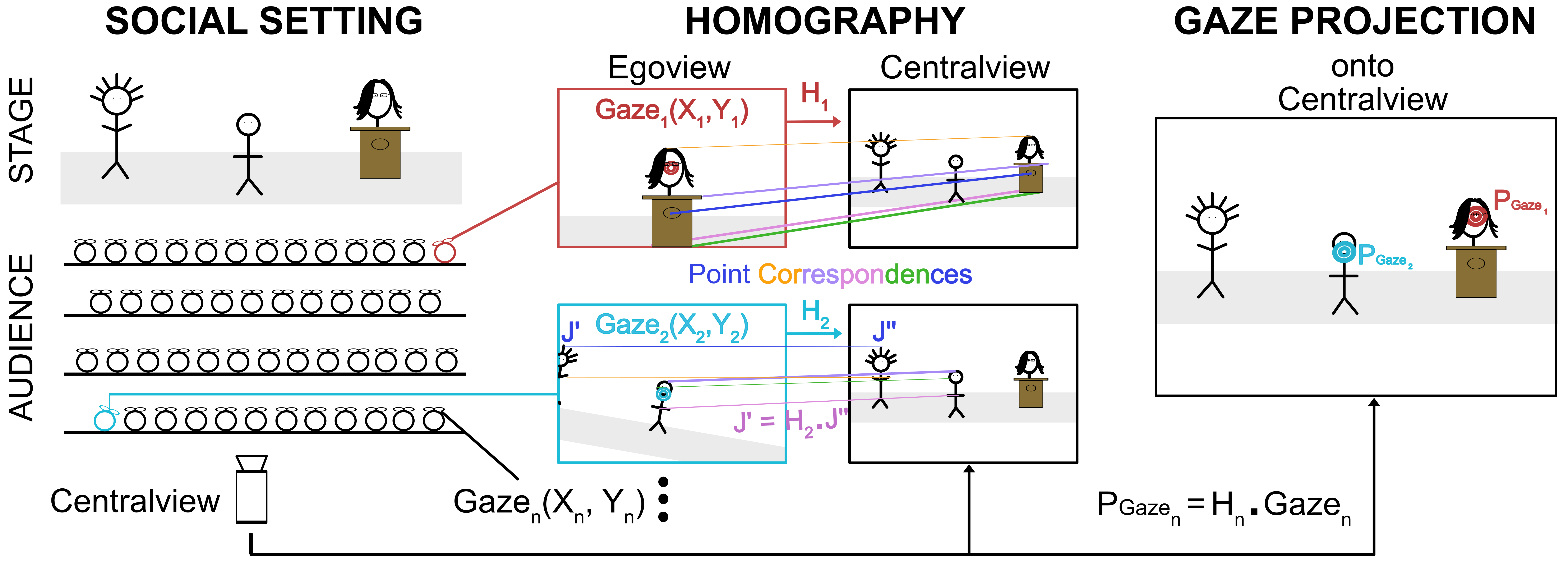}
  \caption{LEFT. Schematic of a shared scene depicting multiple people wearing mobile eye-tracking glasses gazing at a stage. A central camera at the back of the audience records the shared scene (centralview). MIDDLE. {Egoviews} of each glasses wearer need to be projected onto the centralview. Point correspondences between the  egocentric and centralviews are used to calculate homography matrices that relate the two views. RIGHT. Using the computed homography matrices to remap the gaze coordinates, all individuals’ gaze points can be projected onto the shared scene (centralview).}
  \Description[Problem setting]{Figure \ref{problem_setting}. Fully described in the text.}
  \label{problem_setting}
\end{figure}

A schematic of our approach to solve the problem of multi-person eye-tracking in such collective viewing experiences is detailed in Figure \ref{problem_setting}. We imagine a scenario where multiple people observe a shared scene from different viewing angles. The eye movements of each viewer are recorded independently from their egocentric perspective ({egoview}) using commercial-grade eye-tracking glasses. However, to study collective gaze patterns, the wearer-specific gaze data need to be synchronised and mapped onto a shared coordinate space (semantic gaze mapping). We record a centralview of the shared scene from a camera placed behind all observers. We employ Network Time Protocol (NTP) to temporally synchronise data from all devices. The synchronised gaze data from all viewers is then mapped to the shared coordinate space i.e. the centralview by finding robust feature matches between {egoview}--centralview image pairs and performing a projective transformation. Following our approach we design a software framework--SocialEyes. We implement and test SocialEyes during two public events in which we recorded eye data from 30 participants simultaneously. We detail our choices of eye-tracking hardware, the underlying infrastructure, and the implementation specifics of the framework. In validating our approach, we provide visualisation and analysis metrics for multi-person eye-tracking data. \textbf{Overall, we make the following contributions in this paper}:

\aptLtoX{\begin{enumerate}
\item Develop and validate a scalable multi-person eye-tracking system, with flexible modes of operation. Our system provides users the means to monitor incoming data in real-time and to remotely control all associated mobile eye-tracking devices in parallel. We tested our system with 30 mobile eye-tracking devices in parallel, limited only by the number of devices we own. Our system should easily handle more devices when operating in recording mode. This tool opens unprecedented avenues for batch data collection and real-time interactivity in social settings. 
 
\item Solve the problem of multi-perspective gaze data, in the context of shared social scenes. Using our proposed homography approach, it is possible to project the gaze of all eye-tracking glasses wearers from egocentric coordinates to shared central coordinates. This technique revolutionises possibilities for automating analyses of social gaze dynamics in real-world settings. 

\item Introduce and implement standard and new visualisation and analysis techniques for multi-person eye-tracking data. We provide solutions for visualising both egocentric and homography-transformed gaze data and heatmaps, with methods to visualise all individually or in parallel. The transformed gaze heatmaps offer particular promise in analysing group attention and perception. Similarly, analysis metrics like normalised contour area may offer novel insights into collective gaze behaviour.  

\end{enumerate}}{\begin{enumerate}[leftmargin=\dimexpr\parindent+0.7em\relax]
\item Develop and validate a scalable multi-person eye-tracking system, with flexible modes of operation. Our system provides users the means to monitor incoming data in real-time and to remotely control all associated mobile eye-tracking devices in parallel. We tested our system with 30 mobile eye-tracking devices in parallel, limited only by the number of devices we own. Our system should easily handle more devices when operating in recording mode. This tool opens unprecedented avenues for batch data collection and real-time interactivity in social settings. 
 
\item Solve the problem of multi-perspective gaze data, in the context of shared social scenes. Using our proposed homography approach, it is possible to project the gaze of all eye-tracking glasses wearers from egocentric coordinates to shared central coordinates. This technique revolutionises possibilities for automating analyses of social gaze dynamics in real-world settings. 

\item Introduce and implement standard and new visualisation and analysis techniques for multi-person eye-tracking data. We provide solutions for visualising both egocentric and homography-transformed gaze data and heatmaps, with methods to visualise all individually or in parallel. The transformed gaze heatmaps offer particular promise in analysing group attention and perception. Similarly, analysis metrics like normalised contour area may offer novel insights into collective gaze behaviour.  

\end{enumerate}}

\section{Related Work}

Previous work on multi-person eye-tracking has relied on smaller groups (predominantly dyads) and elaborate experimental setups that hinder natural behaviour \cite{bise2024joint, fasold2021gaze, macdonald2018gaze, kera2016, jermann2012duet, kawase2009exploratory, yamada2014visual}. Most studies present static scenes on a screen as a proxy to predict real-world eye-tracking behaviour which was identified as an inefficient approach by \citet{foulsham2017fixations}. Some studies have applied head pose as a proxy for gaze in multi-person setups \cite{stiefelhagen2002tracking, beyan2018}; however, existing research also shows that head orientation is not always correlated with gaze in social-group interactions \cite{vrzakova2016speakers}.  

In addition to the experimental setup, analysis of multi-person eye-tracking data remains a significant hurdle in conducting large scale studies. Previous studies have heavily relied on manual techniques like incorporating gaze reports in surveys \cite{pennill2017rehearsal}, mapping gaze to visual stimuli using proprietary closed-source manufacturer software, or hand-coding regions of interest \cite{benjamins2018gazecode, geeves2014performative}. These techniques are incredibly time-consuming, with poor scalability to large datasets. Further, they are not entirely reproducible, and are prone to human errors. A common strategy applied to simplify the problem is to calculate coarse gaze measures, such as discrete binary classification of gaze on regions of interest \cite{muller2018robust, fasold2021gaze}. While more scalable, such methods provide limited-resolution data, thereby reducing the cost-effectiveness of experiments and limiting the insights that can be gained.

More scalable and automated tools for unrestricted eye-tracking are actively being developed under the umbrella field of ‘pervasive eye-tracking’ \cite{kasneci2017towards}. These are motivated by advances in portable head-mounted eye trackers that have recently become more affordable and accurate. The increased availability of open-sourced and automated analysis tools \cite{panetta2019software, li2006openeyes} has boosted the adoption and potential applications of these mobile eye trackers, which are increasingly being employed in unrestricted natural environments with static scenes, such as art galleries \cite{reitstatter2020display}. However, to date, a synchronised multi-person eye-tracking study in dynamic real world social settings has not been done and consequently the analysis methods to evaluate such large-scale data are not readily available. To solve these challenges we design and implement SocialEyes. 

\section{SocialEyes: Design Concept}

\begin{figure*}[htbp]
  \centering
  \includegraphics[width=0.9\linewidth]{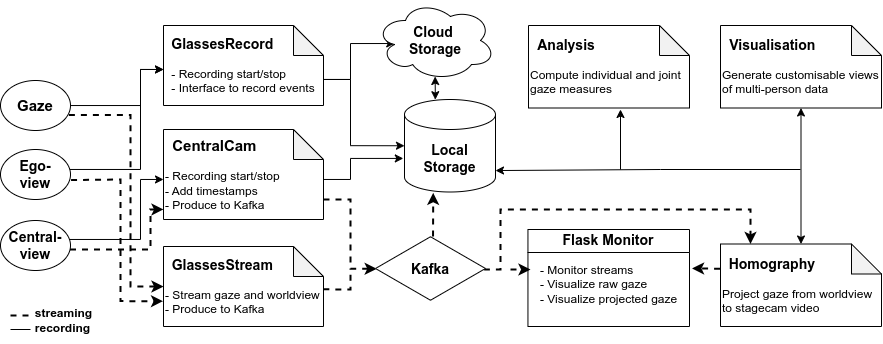}
  \caption{Software framework demonstrating the flow (arrow lines) of data streams (ellipses) in the system. Solid and dashed lines represent the recording and streaming mode of operation respectively. Each mode of operation engages separate modules (dog-eared rectangles) that process the incoming data.}
  \Description[System Design]{Figure \ref{modules}. In recording mode, gaze and egoview data are captured by the GlassesRecord module, while the CentralCam module records centralview data. The data is stored locally or in the cloud for retrieval by the Analysis, Visualization, and Homography modules. In streaming mode, the GlassesStream module records glasses data, which is transmitted via Apache Kafka to subsequent modules.}
  \label{modules}
\end{figure*}

 SocialEyes is a software framework that enables synchronous operation and analysis of multiple eye-tracking data streams in unrestricted, real-world settings. We followed the Research through Design approach \cite{researchThroughDesign} to design a software framework that is independent of specific hardware, network, and storage components, allowing it to adapt to various infrastructure constraints. In our software framework, presented in Figure \ref{modules},  incoming data consists of three parallel streams: i) gaze coordinates from each eye-tracking device ii) {egoview} video recorded from a head-mounted camera on each device, and iii) a centralview video of the shared scene, captured and streamed by a stationary camera mounted to the centre of the scene. Incoming data streams are handled by separate modules that engage based on the mode of operation. Modules in our framework abstract information and allow isolated development and updates of separate components. Below, we describe the different modes of operation.

\subsection{Modes of Operation} \label{Modes of operation}

\subsubsection{Recording} \label{Recording}
The recording mode is designed for offline operation of the framework where data are collected and stored for post-hoc analysis. Such operation is desirable when real-time processing of incoming data is not required, e.g., in research and user studies. Splitting the two stages of data collection and processing reduces network bandwidth and computation requirements, thereby allowing complex analysis and long-duration recordings with limited resources.

\subsubsection{Streaming} \label{Streaming} 
The streaming mode is designed for real-time applications where data are collected and processed concurrently. Processing data in real-time is desirable for interactive applications such as gaze-contingent interfaces. This mode of operation demands sophisticated hardware and software optimisations to achieve good performance. Streaming of data in this mode is facilitated by the open-source event streaming platform--Apache Kafka--providing high-throughput and low-latency. Data streams are processed by Kafka topics, each comprising of multiple partitions that store data in background Java programs called brokers while ensuring load balancing and scalability. Finally, separate consumer modules read these data streams for a) calculating homography matrices, b) writing to local storage, and/or c) publishing the data to a dedicated Flask server for real-time monitoring.

\subsubsection{Recording and Streaming} \label{Recording and streaming}
The recording and streaming mode of operation records raw data concurrently with real-time processing of incoming data streams. This mode provides the flexibility to perform both real-time and post-hoc analysis, at the expense of added computational requirements. Such operation is desirable when applications have gaze-contingent aspects that should be processed in real-time but more complex analyses need to be performed offline.

\subsection{Technical Design} \label{Modules}

\subsubsection{Time-synchronisation}
A critical challenge in managing multiple devices on a network is time synchronization, as each device relies on its internal clock to timestamp data samples. Independent internal clocks must be synchronised to a master clock to ensure accurate timestamp comparisons. All eye-tracking devices, recording cameras, and internal servers are, therefore, synchronised to a master NTP server accessible on the local network. This synchronisation ensures negligible clock offsets between devices, however, internal clocks on these devices can still accumulate variable drifts over long duration recordings. In the streaming mode of operation, this drift can be ignored for most applications to favour low-latency operation. However, for applications requiring fine-grained time alignment (typically in recording mode), we apply the clock-filter algorithm in NTP \cite{mills1994ntpClockFilter} to collect offset and drift at periodic intervals. Offsets between a reference device--the server running SocialEyes software--and each remote device (e.g., eye tracking glasses) are measured by exchanging a brief sequence of User Datagram Protocol (UDP) packets between the two involved devices. Each exchange returns a) an estimate of the round-trip-time (RTT) for the packet, and b) an estimate of the clock offset between the two devices at the time of the measurement, with the round-trip-time factored out. The final offset estimate is the one offset value for which the associated RTT value was minimal. Finally to align the timestamps from a device, its offset information is aggregated over the entire span of the recording (in recording mode) or a fixed size window (in streaming mode). An outlier-robust linear fit is calculated over the aggregated offset measurements using random sample consensus \cite{fischler1981RANSAC} to map the timestamps with the reference device. 
In our design, offset recording is performed as a background process by the GlassesRecord and GlassesStream modules for the recording and streaming mode respectively. The final synchronisation and linear mapping is performed as an optional process in each of the consumer modules i.e. Analysis, Visualisation, and Homography. We detail the functions of these modules below.

\subsubsection{GlassesRecord} \label{GlassesRecord}
The GlassesRecord module is engaged in the recording mode to send recording start and stop signals to each pair of eye-tracking glasses. The module features an intuitive and reliable operator interface to efficiently trigger and monitor actions on multiple devices simultaneously (see example implementation in Section \ref{Utility Test}). Actions can be triggered selectively on a subset of devices to ensure fault tolerance, for example, the ability to restart recording on one pair of glasses if it fails, rather than having to restart on all devices. The module also provides functionality to annotate recordings with custom hand-marked events.

\subsubsection{CentralCam} \label{CentralCam}
The CentralCam module manages a centrally mounted camera that captures a stationary view of the shared scene. This camera can be interfaced as a USB peripheral or a network device via Real-Time Streaming Protocol (RTSP). The module grabs and timestamps independent frames from the incoming camera stream. It also computes peformance metrics such as dynamic frame-rate in frames per second (FPS), temporal jitter calculated as the mean and standard deviation of inter-frame time difference, and dropped frame counts.  In recording mode, the module offers flexibility to store the recorded video in an mp4 container or a raw format. The raw format, while consuming more storage space, requires less computational overhead since it avoids decoding and preserves lossless data for later processing or correction. In the streaming mode, the module initialises a Kafka producer and pushes each incoming frame and its timestamp to the producer. 
 
\subsubsection{GlassesStream} \label{GlassesStream} 
GlassStream implements the streaming of gaze and {egoview} data streams in the streaming mode of operation. The module initialises a Kafka producer, parses incoming gaze and video data streams from selected eye-tracking devices on the network and publishes them to the Kafka cluster. The module also initiates a metrics server, incorporating Grafana and Prometheus services, to monitor the Kafka cluster.

\subsubsection{Flask Monitor} \label{Flask Monitor} 
The Flask Monitor module provides a real-time display of the incoming data streams to verify data integrity. The module initialises a Kafka consumer that reads each topic and updates corresponding graphs for that data stream. The module hosts a flask web-application that displays {egoview} streams from all devices with a configurable refresh rate (default 1Hz) and incoming frame rate (in FPS). Incoming gaze is displayed as a dynamically updating time-series of x and y coordinates, with a configurable refresh rate. The sample rates for gaze and {egoview} streams are set to a low value by default to reduce stress on the server and keep the monitor interface responsive. The incoming centralview is consumed in a separate process and displayed in a separate window, similar to the {egoview} frames. The module interface also features alert mechanisms to notify when data flow ceases or the frame rate drops.

\subsubsection{Homography} \label{Homography}

The homography module implements semantic gaze mapping from {egoview} coordinate space to centralview coordinate space simultaneously for multiple devices (see Figure \ref{problem_setting}). Applying a pin-hole camera model, the {egoview} captured from each wearer's perspective can be assumed to lie on a planar surface at an infinite distance. The projective transformation between the {egoview} and centralview for the same scene is then mathematically related by a planar homography. Incoming gaze samples (x, y coordinates) and {egoview} frames from each device are paired together with the nearest centralview frame using the marked timestamps from the respective input modules of each data stream. The module accommodates varying data sampling rates of the three data streams, since each stream's sampling resolution depends upon factors like hardware configuration, recording parameters, and network bandwidth. The final temporal resolution of the homography module output is the same as the sampling rate of the provided centralview (i.e., for each centralview frame, the homography module attempts to map all gaze points from the provided {egoviews}+gaze pairs to that centralview frame). This mapping is accomplished by finding robust feature point correspondences between each {egoview}-centralview image pair. Matched key points are then used to calculate a homography matrix that is finally used to transform gaze from {egoview} to centralview coordinate space. 

The performance of this module depends significantly on detecting and tracking robust feature point correspondences in dynamic, real-world social settings which often involve challenging artefacts caused by illumination changes, motion distortion, low-textured regions, and occlusions. In recording mode, the module can utilize large deep learning models, such as the state-of-the-art SuperGlue \cite{sarlin2020superglue} model demonstrated in our Utility Test (see Section \ref{Utility Test}), thanks to the absence of computational and time constraints. However, for real-time operation in streaming mode, simpler algorithms like \citet{bay2008SURF, rublee2011ORB, tomasi1991LKT} are preferred, balancing the trade-off between speed and accuracy.

\subsubsection{Visualisation} \label{Visualisation}
The visualisation module aligns timestamped {egoview} frames, gaze samples, and centralview frames with an additional input of the transformed gaze (in the centralview space) computed by the homography module. We design the following visualisation modes in this module:

\noindentparagraph{\textbf{Single-person gaze and egoview}:} Overlays gaze synchronised to the {egoview} stream for each selected viewer as a separate video file with an optional overlay of associated timestamps for each data sample.
\noindentparagraph{\textbf{Multi-person gaze and egoview}:} Renders a single video with all selected viewers’ {egoviews} with the corresponding gaze points overlay (as in the previous option) in a 2D grid of customisable size.  
\noindentparagraph{\textbf{Single-person egocentric and transformed gaze view}:} Stacks the time-synchronised {egoview} and centralview frames for a single viewer along with the corresponding gaze and transformed gaze overlay on each view respectively.
\noindentparagraph{\textbf{Multi-person transformed gaze on centralview}:} Presents the centralview with all time synchronised gaze coordinates from the selected viewers overlaid on top. The transformed gaze for each viewer can be plotted either as customisable circles containing a unique ID and outline colour, or as a heatmap overlay calculated from the 2D distribution of all transformed gaze points.
\noindentparagraph{\textbf{Multi-person egocentric views with transformed gaze on centralview}:} Combines the {egoviews} from all selected viewers and plots it as a grid with the time synchronised centralview in the centre cell. Gaze for each viewer is overlaid in their respective {egoview} and the centre cell displays all the transformed gaze points similar to the previous option.

\subsubsection{Analysis} \label{Analysis}
The Analysis module provides methods to preprocess and evaluate the recorded egocentric gaze data as well as the homography-transformed gaze data for collective gaze analysis. It includes the following metrics combining traditional eye-tracking measures, such as heatmap similarity and correlation, scaled up to multi-person use cases, with novel collective measures, such as normalised contour area, to quantify gaze dynamics in social group settings:

\noindentparagraph{\textbf{Average Gaze Velocity}:} Gaze velocity is calculated as a time series of inter-sample Euclidean distances of gaze (x, y coordinates) samples. The average gaze velocity over a given timeframe quantifies mean precision of the gaze predictions which reflects overall reliability of the gaze predictions.
\noindentparagraph{\textbf{Heatmap}:} Gaze heatmaps are calculated from a series of x and y gaze coordinates by convolving a 2D Gaussian filter with sigma (standard deviation of the gaussian kernel) equal to the approximate size of the fovea (i.e., one degree of visual angle). 
\noindentparagraph{\textbf{Heatmap Similarity (SIM) and Correlation (CC)}:} SIM is calculated as the histogram intersection between the distributions of two heatmaps (ranges between 0 and 1 where 1 represents identical distributions) and CC is represented by the Pearson correlation coefficient between them to quantify the degree of linear relationship between their pixel intensities.
\noindentparagraph{\textbf{Stationary Gaze Entropy}:} Stationary gaze entropy is calculated by organising gaze coordinates in a given time frame into spatial bins of roughly one degree of visual angle (after conversion to the same coordinate space). Shannon’s entropy equation is then applied to this discrete probability distribution of gaze to calculate the average level of uncertainty or the overall predictability of the gaze coordinates. The obtained entropy quantifies gaze dispersion in the given time frame with a higher entropy or uncertainty indicating a wider distribution of gaze across the visual field \cite{GazeEntropy}.
\noindentparagraph{\textbf{Normalised Contour Area}:} Normalised contour area provides a dynamic measure of gaze dispersion over time. It is calculated as the area of the smallest convex polygon that encloses all provided gaze samples at a given time point. The convex hull was initially adapted to eye-tracking analysis by Goldberg and Kotval \cite{ConvexHull}.

\section{Utility Test} \label{Utility Test}

To test the viability and scalability of our framework with a large audience in a real-world context, we
implemented the framework during a public event featuring a musical concert and a film screening. The event was held at the Large Interactive Virtual Environment (LIVELab) at McMaster University, Canada, which is a custom-built research concert hall with fully customisable spatial audio and room acoustics. The venue holds 106 audience members; we had mobile eye-tracking glasses for 30 audience members. Below, we detail our component choices for the implementation of SocialEyes in recording mode and provide results from the homography, analysis and visualisation modules to derive insights from multi-person eye movement data.

\subsection{Method} \label{Methods}
\subsubsection{Apparatus} \label{Apparatus}

\begin{figure}[htbp]
  \centering
  \includegraphics[width=0.95\linewidth]{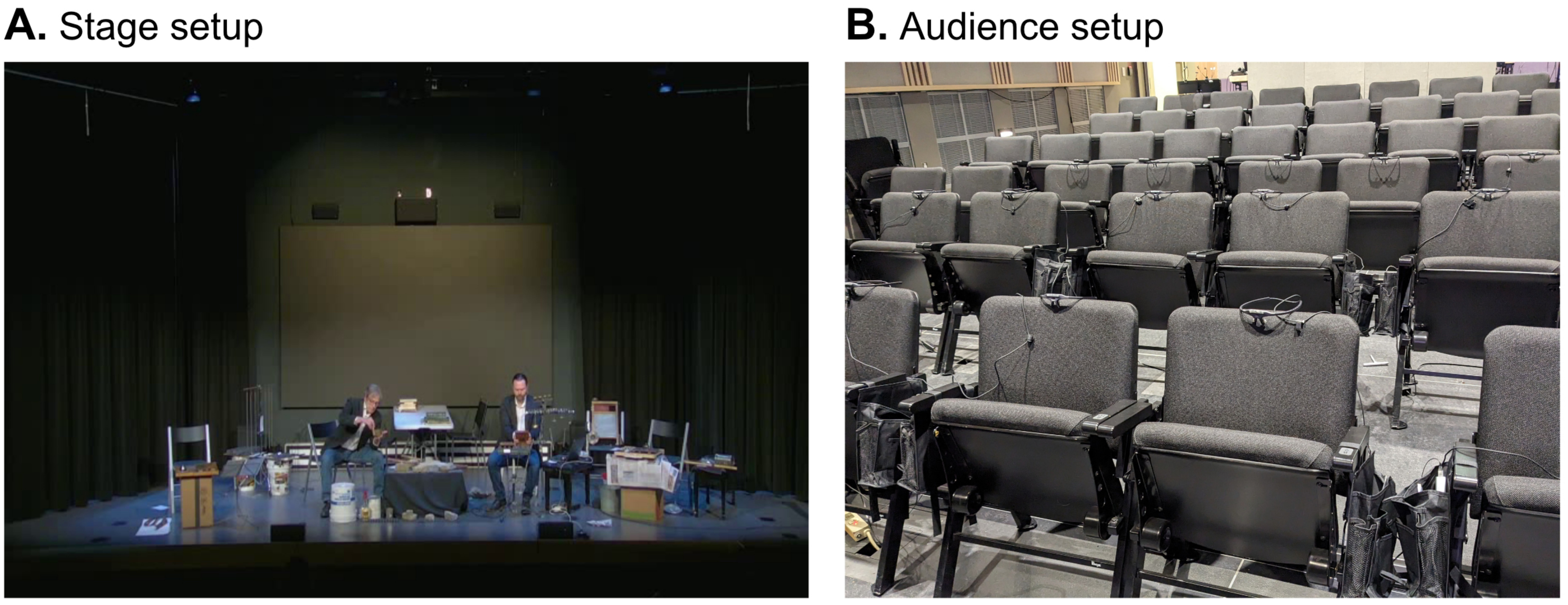}
  \caption{A. represents the stage setup during the concert part of the event. The film was presented on the video wall visible behind the performers. During the film, the stage was cleared and/or covered. B. shows audience chairs with eye-tracking glasses attached to each seat. The eye-tracking glasses were connected to companion smartphones that were secured in pockets attached on either side of each pair of seats. }
  \Description[Study setup]{Figure \ref{setup}. Fully described in the text}
  \label{setup}
\end{figure}

\noindentparagraph{\textbf{Eye-tracking equipment}:} We used the Pupil Labs Neon eye - tracking devices \cite{baumann2023neon} that apply a deep learning pipeline to provide reliable gaze predictions in unrestricted settings. At the core of these eye trackers is a hardware module equipped with binocular infrared cameras that capture eye images with a resolution of 192 x 192 pixels at 200 Hz and a front-facing RGB camera that records the {egoview} with a resolution of 1600 x 1200 pixels at 30 Hz. {The fixed geometry of eye and egoview cameras on the module eliminates the requirement of explicit eye-egoview calibration before each recording; the gaze is recorded in egoview coordinates.} The module is mounted to a 3D-printed, wearable frame that connects to an accompanying Android smartphone (Motorola Edge 40 Pro). The minimal form factor of the frames is designed to be unobtrusive and effortless for participants, making them a good fit for longer-duration studies. The accompanying Android phone runs a proprietary vendor application–“Neon Companion” (v2.8.2) –that hosts a suite of HTTP REST Application Programming Interfaces (APIs) enabling remote streaming of data and controlling operations such as start/stop recording. 
During the data collection event, each pair of eye-tracking glasses was connected to the companion smartphone device running Pupil’s Labs NEON Companion android application in the background. The phones were locked and connected to an Anker 6-in-1 USB-C hub that extended the ports to connect them to ethernet, power, and the NEON glasses concurrently. Each smartphone was secured in a bag attached to the respective participant’s chair, ensuring free movement of the seated participant (see Figure \ref{setup}B). The ethernet cable connected the device to a network switch placed under the false floor, in the centre of the audience rows. The global NTP server for all smartphones, the central camera, and the server computers were all set to the University’s NTP server accessible on the local network.

\noindentparagraph{\textbf{Centralview camera}:} A PTZOptics pt30x - sdi - g2 camera was mounted central to the stage, positioned behind the audience seating. The camera captured an unobstructed view of the stage with a resolution of 720x480 px at 60 Hz. The camera feed was streamed on the network via RTSP with H264 encoding. Figure \ref{setup}A demonstrates an example frame from the central camera. 
\noindentparagraph{\textbf{Presentation screen}:} The film was presented on a 14.5 ft (384.5 cm x 216 cm) Samsung LH015IER LED cabinet with a resolution of 1920x1080 pixels at a 60 Hz refresh rate and a pixel pitch of 1.5 mm. 
\noindentparagraph{\textbf{Computing servers}:} The GlassesRecord and CentralCam modules were executed on separate desktop servers running Ubuntu 22.04 LTS Linux distribution. Note that both modules have low CPU requirements and could be easily executed on a single general-purpose server, as long as sufficient storage is available for long-duration recordings. We leveraged the modular structure of our framework to run the modules on separate servers to allow separate operators and monitors for the two modules.
\noindentparagraph{\textbf{Network infrastructure}:} Each pair of Neon glasses requires a bandwidth of 6-8 Megabytes per second (MBps) to stream gaze and {egoview} data. For real-time streaming operations, it is critical to consider the available bandwidth on a network to identify how many concurrent eye-tracking glasses’ data can be streamed. In this test, we used the recording mode of operation which has lower bandwidth requirements, since real-time data streams do not need to be accessed on the server. Nonetheless, we leveraged the stationary positioning of participants in our setup to provide precise timing information and optimal network reliability by connecting each eye-tracking device to a network switch (TP Link TL-SG1048) via ethernet. The network did not have internet access, to disable any background network activity on the connected Android smartphones.
\noindentparagraph{\textbf{Storage}:} {egoview} videos and gaze data were recorded and stored locally on each smartphone through the Neon companion app. The recorded data was later uploaded to Pupil Cloud for post-processing of the recorded raw data and then downloaded for further analysis. Time offsets and logs from the GlassesRecord module were stored locally on the server running the module. Centralview video and metrics recorded by the CentralCam module were stored locally on the server running the CentralCam module.

\subsubsection{SocialEyes implementation in Recording mode} \label{SocialEyes implementation}


\noindentparagraph{\textbf{GlassesRecord}:} We implemented the GlassesRecord module utilising the pupil-labs-real-time-api python library (\href{https://pupil-labs-realtime-api.readthedocs.io}{https://pupil-labs-realtime-api.readthedocs.io}) to send start, stop, discard, and save recording triggers to each Neon device. Periodic clock offset measurements were recorded every ten seconds. Additionally, since the accompanying smartphones use Android as operating system, we used Android Debug Bridge (ADB) to retrieve status metrics including battery level, storage level, network ping, ADB server status, and connected USB devices. The status metrics facilitated continuous monitoring of devices and recording status. ADB was also used to remotely control the Android smartphones for remote troubleshooting and verifying indicators of failed or corrupted recordings that were not reported by the vendor application. These operations were important to prevent data loss and ensure that we would never need to interrupt the live social event to access a participants’ phone in the event of an error. The operator interface for this module was implemented as a Terminal User Interface (TUI) built using the python Textual library (\href{https://textual.textualize.io/}{https://textual.textualize.io/}), providing low runtime requirements and cross platform operation. {The TUI displays a list of all eye-tracking devices on the local network, along with colour-coded status metrics for each device (battery/storage level, network ping, ADB server status, recording status, etc.). Multiple devices can be selected from this list to perform a desirable action with dedicated keystrokes. The following actions can be performed on all selected devices using the TUI: Start Recording; Save Recording (end and save recordings locally); Cancel Recording (end and discard ongoing recordings); Restart App (use ADB to restart the Neon Companion app, critical for troubleshooting app crashes remotely); Reconnect ADB (reconnect to wireless ADB, critical for recovering lost connection to devices).}


\noindentparagraph{\textbf{CentralCam}:} The CentralCam module was implemented with OpenCV \cite{opencv_library} and FFMPEG (\href{https://ffmpeg.org}{https://ffmpeg.org}) libraries to retrieve the RTSP stream from the central camera. The incoming stream was compressed to a MPEG4 container using an H264 codec. A tolerance of 600 dropped frames was added to ignore intermittent packet drops on the network before the module terminates recording. FPS was calculated over a window of 180 frames and stored with other metrics to a CSV file format.
\noindentparagraph{\textbf{Homography, Analysis, and Visualisation}:} We applied a pre-trained Graph Neural Network–-SuperPoint/ SuperGlue \cite{sarlin2020superglue}--to find robust feature point correspondences in our Homography module. Blinks were identified using the Neon blink-detection algorithm {\cite{pupil-blinks}}
 and filtered out of the raw gaze data before homography, analysis and visualisation. The operator interface for these modules was implemented as a command-line utility using the Python questionary (\href{questionary.readthedocs.io}{questionary.readthedocs.io}) library. The command-line utility offered formatting of the directory structure for compatibility with modules, selecting a subset of recordings to perform functions, such as homography, visualisation, gaze metrics calculation, etc., and storing output to the local filesystem. {Gaze heatmaps were generated using the viridis scientific colormap \cite{viridis}, which is designed for accurate perceptual uniformity and accessibility for viewers with color vision deficiencies.}
\noindentparagraph{\textbf{Visual angle calculation}:} Visual angle conversion was done by estimating the distances of each of the three participant rows from the videowall. The distances were 758 cm for the first, 885 cm for the second, and 1012 cm for the third participant row. Since the physical dimensions of the video wall were 384.5 cm for width and 216 cm for height, the diagonal (441.02 cm) projected a visual angle of 32.4° for the first, 28° for the second, and 25.6° for the third row. From the {egoview} recordings of participants in each row, the diagonal of the videowall was estimated to be 515 pixels for the first, 441 pixels for the second, and 396 pixels for the third row. The visual angle to pixel ratio was therefore calculated to be approximately 16 (i.e. one visual degree was represented by roughly 16 pixels in the 1600x1200 resolution {egoview} recordings).

\subsubsection{Stimuli} \label{Stimuli} 


\noindentparagraph{\textbf{Music Performance}:} {The one hour long composition consisted of home\-made instruments, spoken texts, and electronic soundscapes to present the audience with dramatic soundscapes that explore issues related to the American criminal justice system. Of the 17 total tableaux in the performance, some were too long, uncomfortable, confusing; others were direct, melodic, and in familiar styles. }
\noindentparagraph{\textbf{Film}:}{The film showcases the compositional process behind the performance, engaging with exonerees, whose story and reaction to the performance is also captured.}

\subsubsection{Study Design}
The current utility test takes place within the context of a larger project examining the role of context information in altering audience physiological responses and social attitudes and behaviours. During the event, in addition to wearing eye-tracking glasses, participants also wore a smart watch to monitor cardiac activity and completed a series of surveys. Parallel to this in-person event, we also presented a livestream online. We recorded eye measures, via web browser and webcam \cite{saxena2022towards, saxena2023deep}. We were interested in the effect social co-presence might have on our measures of interest. Detailed analyses related to our cognitive neuroscientific and social psychological questions will be reported elsewhere. In the current study, we focus on validating the utility of our software framework--SocialEyes.

\subsubsection{Procedure} \label{Procedure} 
The study took place over 2 days (April 2nd and 4th, 2024) where the same event was repeated in the opposite order: film before the performance on day one and film after the performance on day two. In the following text, we refer to these as four recording sessions, with the corresponding set of participants in each session as Group1 Film (G1F), Group1 Performance (G1P) and Group2 Performance (G2P), Group2 Film (G2F). The event was open to the general public. Participants were recruited either after purchasing advanced tickets before the event, or upon arrival on-site at each event. The study was approved by the McMaster Research Ethics Board (MREB \#1975) and all participants provided consent before participating. The same procedure was repeated on both days.

Before the event, ADB connection was established between the server and each smartphone device. A forced synchronisation to the assigned NTP server on the smartphones was also done. Participants were seated evenly across rows two to four in the audience, hereby referred to as participant-rows one to three in this paper. After taking their seat, participants were instructed on how to put the glasses on. Trained research assistants completed a quick calibration procedure. {Note that explicit eye-egoview calibration was not needed for the eye-tracking hardware (see Section \ref{Apparatus}). The calibration procedure mentioned here simply performs an individual offset correction using a single fixation target, which can account for physiological factors, such as the Kappa angle between the optical and visual axis of the eye.} Participants retained their seating positions over the entire study duration. 

To ensure proper connection to all eye-tracking devices, test recordings for a short length {(1-5 minutes)} were made before the actual recording session was started. We started the recording session before the event host came on stage. After the start of the event, no interruptions were allowed and all interactions with the recording devices were done remotely using the GlassesRecord module TUI to trigger recordings and monitor the eye-tracking devices (see {GlassesRecord in Section \ref{SocialEyes implementation}}).

In between film and performance (or vice versa), there was an intermission of 30 minutes, during which we stopped recording and participants took off their glasses, completed a short survey, went to the lobby, used the restroom, etc. The central camera recording was not stopped during event intermission. After intermission, the same calibration procedure was repeated before the second half of the event started. Recordings were stopped when the host came onstage at the end of the second half to close the event. The event started at 7 pm and was, on average, 3 hours and 15 minutes long for each of the two days (including intermission). 


\subsubsection{Participants} \label{Participants}
A total of 60 people participated in the study (30 on each day, no repeated participants). Data from one of the participants was only recorded for half of the study due to dropout at intermission. Of the remaining 59 participants, 38 self-identified as women and the reported mean age was 34 years, ranging from 16 to 82 years. All participants had normal or corrected-to-normal vision (via contact lenses; participation while wearing another pair of glasses was not possible). 

\subsection{Results} \label{Results}

\subsubsection{Time Synchronisation} \label{Time Synchronisation}


\begin{figure*}[htbp]
  \centering
  \includegraphics[width=0.8\linewidth]{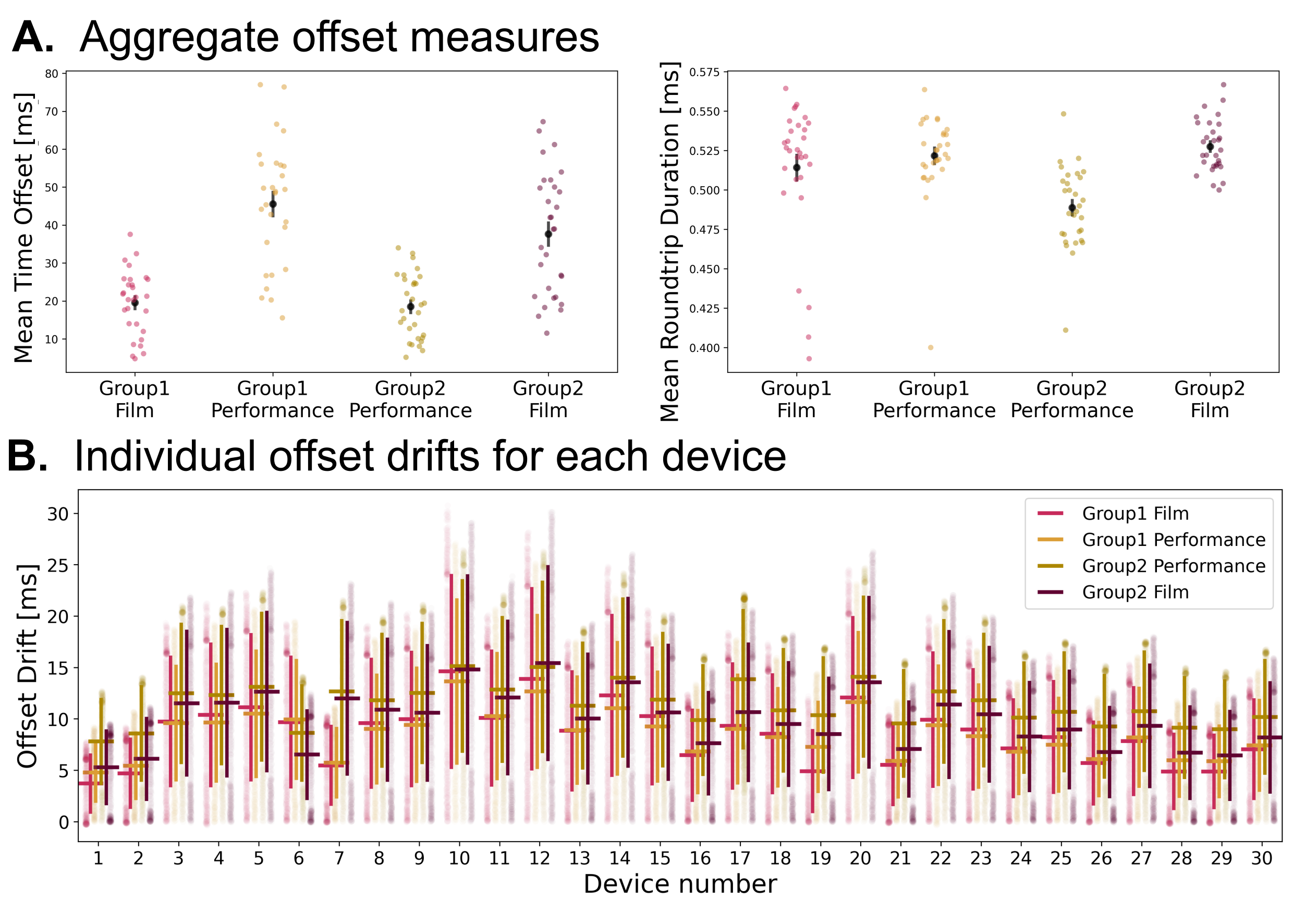}
  \caption{ A. Mean time offset (left) and mean roundtrip duration (right), in ms, recorded for the two sessions on each day. The coloured dots represent mean offset for each device over the respective session (x-axis); black dots represent mean offset across devices with the error bars representing standard error. B. Offset drift calculated for each device (x-axis) in each of the four sessions. The vertical bars represent standard deviation, horizontal bars represent the mean and individual dots represent single data points at measured timepoints.}
  \Description[Time synchronisation results]{Figure \ref{offsets}. A. Mean time offsets are presented on a y-axis range of 0-80 ms. Mean offsets in the second recording session is higher than the first session on both days. Mean Roundtrip durations are presented in the right plot on a y-axis range of 0 - 0.575 ms. B. Drifts in offset for each device is plotted on a y-axis range of 0-30 ms. Spread of the recorded drifts vary slightly between devices but the recorded drifts do not contain spurious outliers.}
  \label{offsets}
\end{figure*}

\noindentparagraph{\textbf{Offsets}:} Time offsets and mean roundtrip duration between each device and the server were recorded at regular time intervals of 10 seconds during each of the four sessions. The mean offset over all devices was 19.58 ms (SD = 8.50 ms) for G1F, 45.57 ms (SD = 16.45 ms) for G1P, 18.52 ms (SD = 8.46 ms) for G2P, and 37.65 ms (SD = 16.14 ms) for G2F; see Figure \ref{offsets}A. Mean offsets were computed after outlier rejection of 3 data points that had high offsets (-277.4 ms, -261.4 ms, and 277.3 ms) due to a failure in the initial time synchronisation procedure for the respective devices. The higher offset of the outliers, however, does not hinder their time synchronisation with other devices since the relative offset as well as the over-time drift of the offset were recorded. The mean roundtrip duration over all 30 devices, was 0.51 ms (SD = 0.04 ms) for G1F, 0.52 ms G1P (SD = 0.03 ms), 0.49 ms (SD = 0.03 ms) for G2P, and 0.53 ms (SD = 0.02 ms) for G2F; see Figure \ref{offsets}B. 
\noindentparagraph{\textbf{Drifts}:} Accumulated drifts in time offsets were calculated as the difference between the offset at a given time point from the initial offset. The mean drift over all 30 devices for G1F was 8.58 ms (SD = 6.48 ms); G1P: 8.48 ms (SD = 5.54 ms); G2P: 11.52 ms (SD = 6.53 ms); G2F: 9.92 ms (SD = 6.88 ms).
\noindentparagraph{\textbf{RANSAC Fit}:} The mean accuracy score of the linear fits of the 30 devices for each session was 0.95 (SD = 0.03) for G1F, 0.95 (SD = 0.04) for G1P, 0.95 (SD = 0.06) for G2P, and 0.95 (SD = 0.04) for G2F.
\par Overall, these data show relatively low offsets and drifts across all 30 devices and recording sessions. As is expected, mean offsets increase from the 1st to 2nd recording session for both groups (Figure \ref{offsets}A; left panel), while drifts seem more so to be device-specific than recording session-specific. Regardless of the actual calculated offsets and drifts, our RANSAC approach allows us to accurately align the timestamps between each eye-tracking mobile device and our central camera.   

\subsubsection{Visualisation}

\begin{figure*}[htbp]
  \centering
  \includegraphics[width=0.85\linewidth]{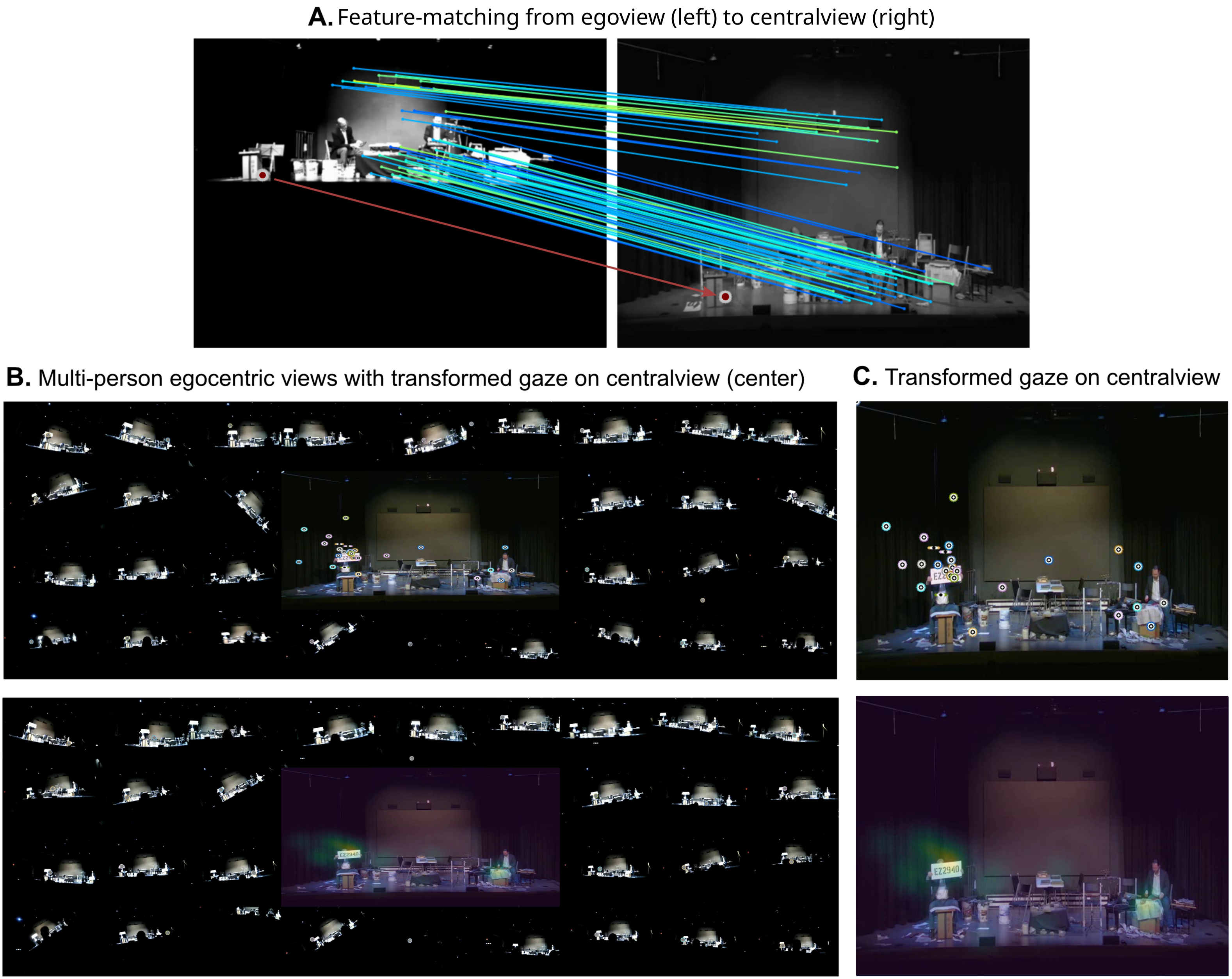}
  \caption{A. An example of feature matching from one frame of one participant’s {egoview} (left) onto the centralview (right). The recorded gaze corresponding to this frame is represented by the maroon and white circle. B. Multi-person views generated with the visualisation module. Top: All participants’ {egoviews} and gazes (small white and black circles) are displayed around the perimeter. SocialEyes projects each participant's gaze onto the centralview (centre). In the centre panel, participants’ gazes are represented with uniquely coloured circles with white in the middle to aid visibility. Bottom: Same as top, with the centre grid cell displaying a heatmap of the 2D gaze density of all participants looking at the scene. Higher intensities in the heatmap (intensity increases from purple to yellow with yellow being the highest) represent a higher proportion of the participants looking at that location. C. Same as centre panels in B, but resized to aid visibility.}
  \Description[Visualisations from SocialEyes]{Figure \ref{viz}. Fully described in the text.}
  \label{viz}
\end{figure*}

\begin{figure*}[htbp]
  \centering
  \includegraphics[width=0.76\linewidth]{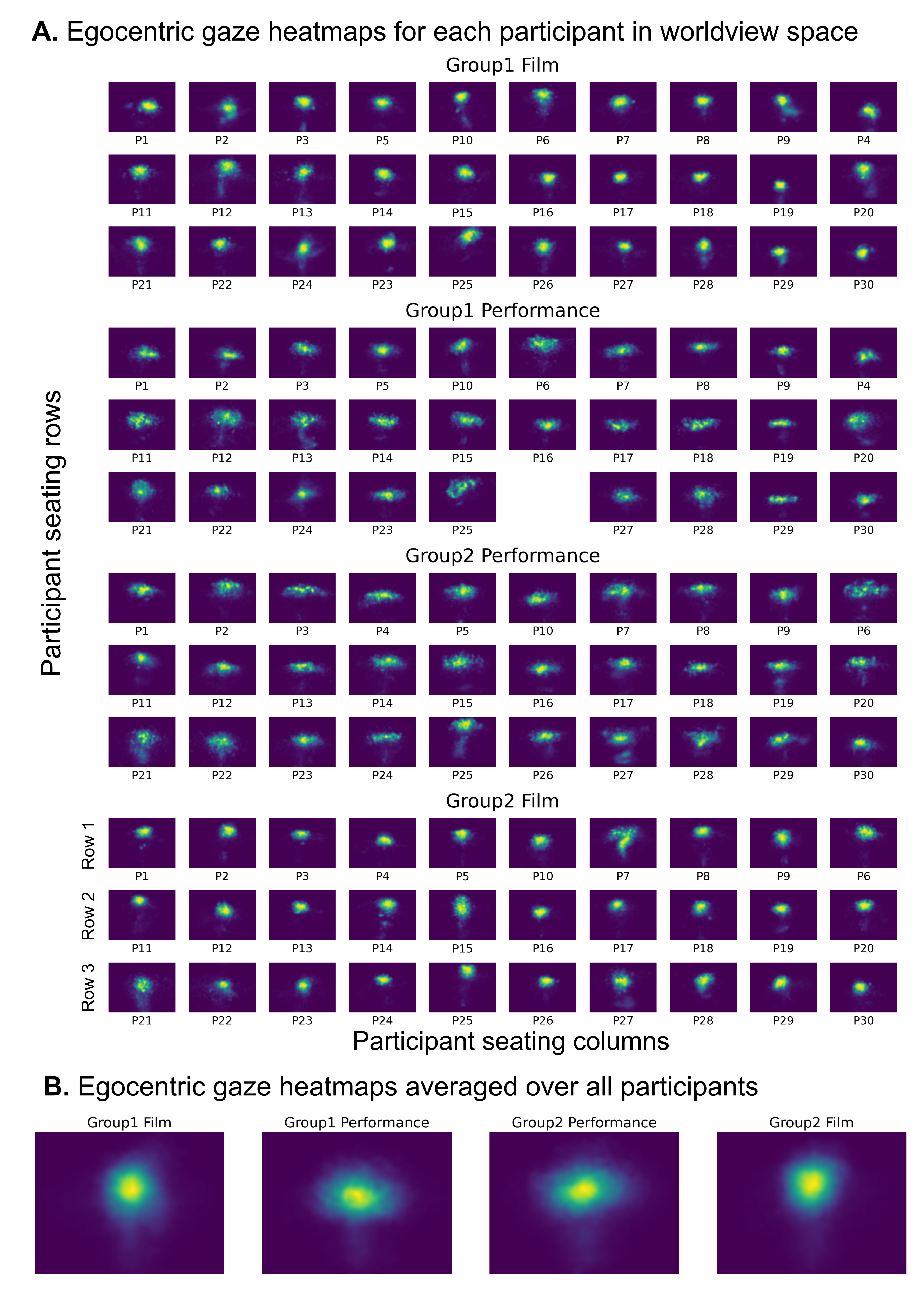}
  \caption{ A. Participant-wise gaze heatmaps for each of the four event sessions. The heatmaps are arranged according to the seating order on both days where the top row (Row 1) is closest to the stage and the first column (from the left) is the leftmost participant. The sessions are arranged chronologically from top to bottom panel. B. Averaged gaze heatmaps calculated as the mean over all participants for each session. }
  \Description[Egocentric gaze heatmaps]{Figure \ref{heatmap_ego}. Fully described in the text.}
  \label{heatmap_ego}
\end{figure*}

\begin{figure*}[htbp]
  \centering
  \includegraphics[width=0.74\linewidth]{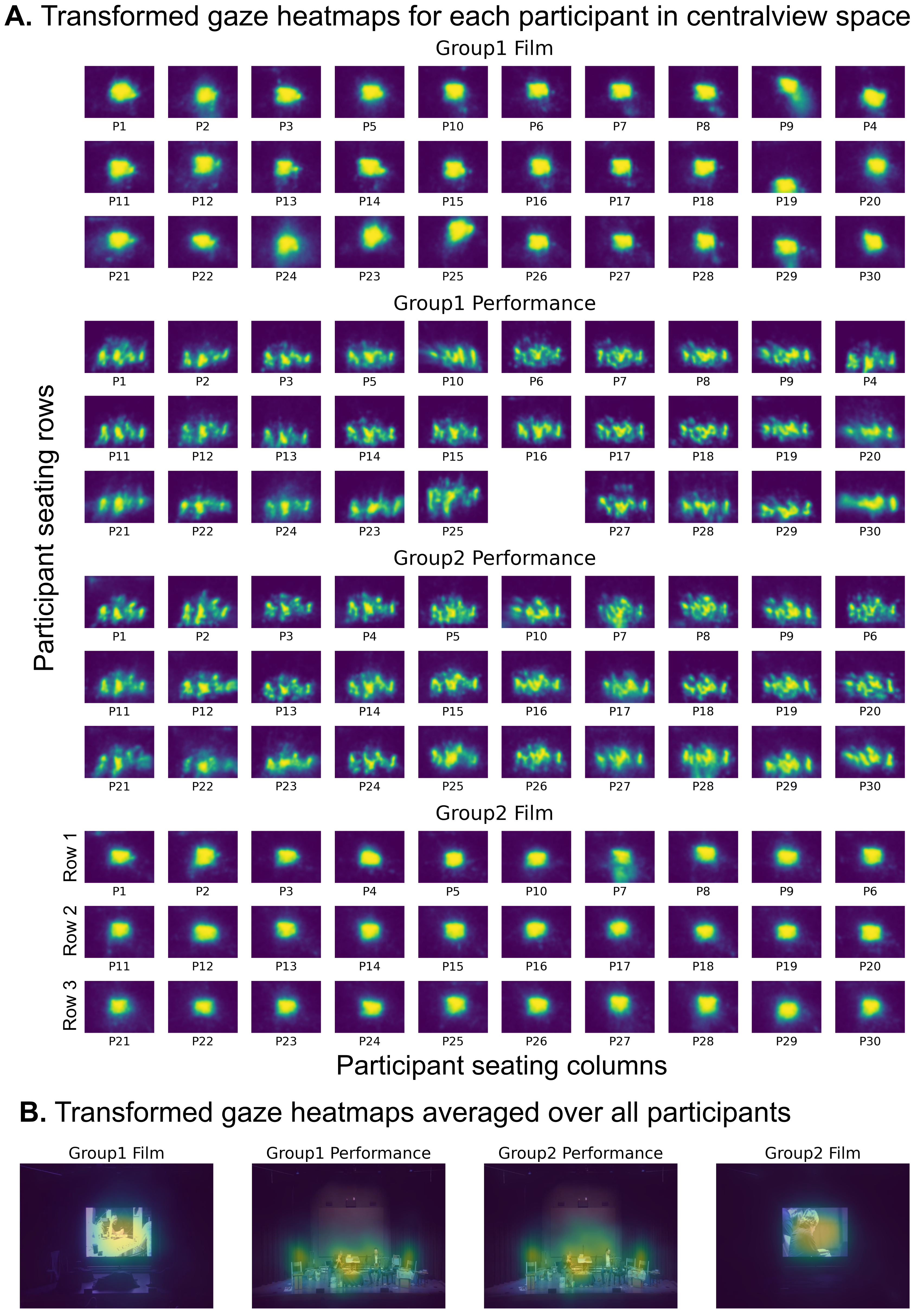}
  \caption{A. Participant-wise gaze heatmaps for each of the event sessions. The heatmaps are arranged according to the seating order on both days where the top row (Row 1) is closest to the stage and the first column (from the left) is the leftmost participant. The sessions are arranged chronologically from top to bottom. B. Averaged gaze heatmaps calculated as the mean over all participants for each session.}
  \Description[Transformed gaze heatmaps]{Figure \ref{heatmap_hom}. Fully described in the text.}
  \label{heatmap_hom}
\end{figure*}

\begin{figure}[htbp]
  \centering
  \includegraphics[width=\linewidth]{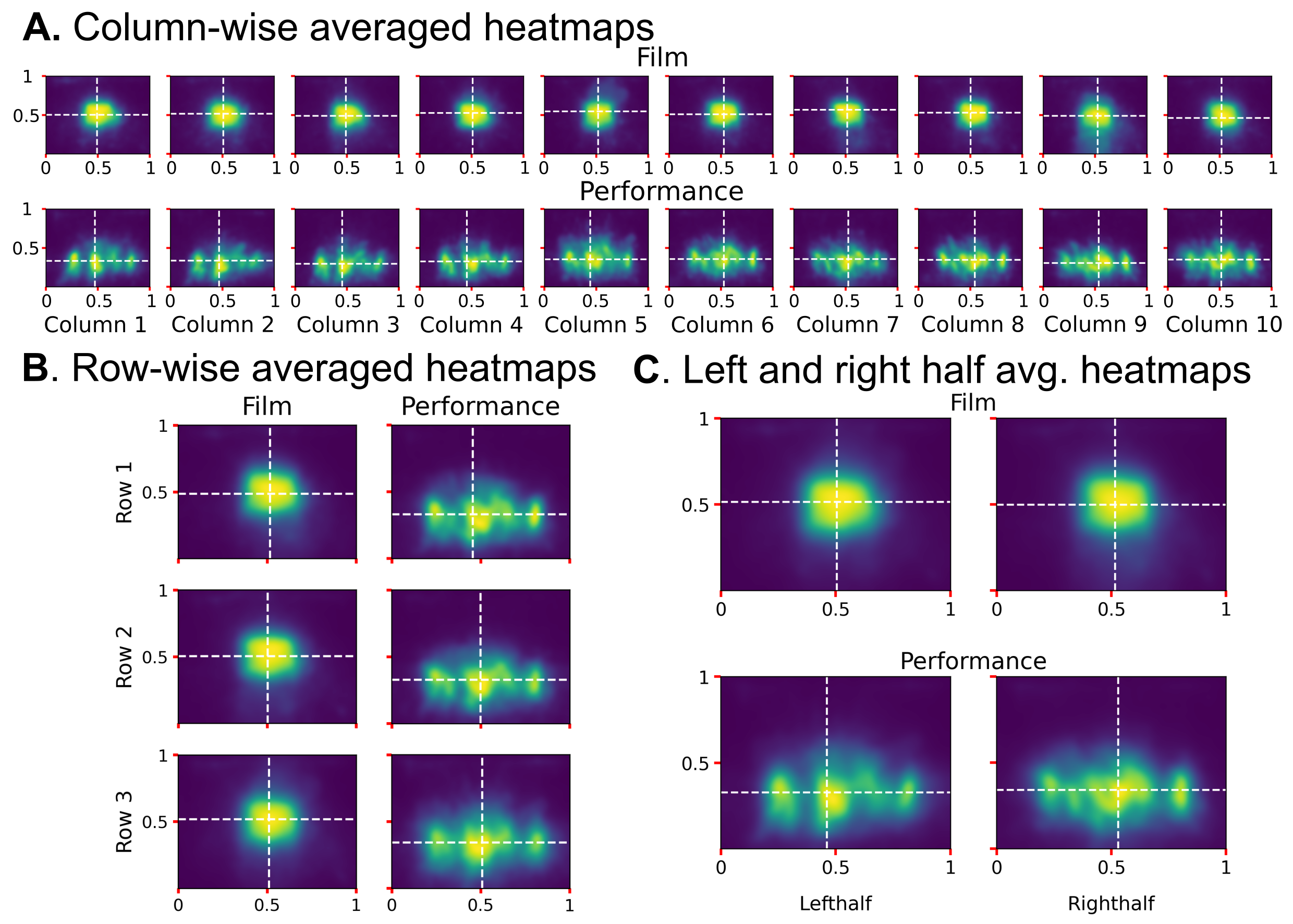}
  \caption{Grouped average of homography-transformed gaze heatmaps. Centres for both x and y coordinate axes are represented by the black ticks on the bottom and left axes respectively. The axes are normalised with the centers marked as 0.5. The white dashed lines represent the location of peak heatmap intensity in the respective coordinate axis. \textbf{A}. shows column-wise averages of all rows over the two event days. \textbf{B}. shows row-wise averages of all columns of the two days. \textbf{C}. shows heatmap averages for the left (left column) and right (right column) halves of the audience (averaging all rows for columns 1-5 and 6-10 respectively). }
  \Description[Averaged heatmaps]{Figure \ref{heatmap_avg}. Fully described in the text.}
  \label{heatmap_avg}
\end{figure}

Egocentric and homography-transformed gaze coordinates were projected onto each viewer’s {egoview} and the shared centralview, respectively, after removal of blinks.  The visualisations (presented in Figure \ref{viz}) enable data validation and demonstrate accurate projective transformation using the SuperGlue features identified in the Homography module (see Section \ref{Homography}). The transforms {computed using our SuperGlue + RANSAC approach} handle video artefacts, such as motion blur (from sudden head movements), poor lighting/texture (frames with large dark regions, shadows, lens flare, glare etc.), obstructions (audience members in previous rows), and scale variations (different resolution and zoom in centralview and {egoviews}) reliably. However, further optimisations and investigations would be needed to identify edge cases in the homography estimation {(we discuss these optimisations in Section \ref{edge-cases})}. Figure \ref{viz} displays still, single frame examples of our transformed data; an example video can be viewed here: \href{https://tinyurl.com/socialeyes-beatlab}{https://tinyurl.com/socialeyes-beatlab}. In the first part of this video example (tableau: “canjo”), the two performers are seated very near to each other. Even at this close distance, we can clearly distinguish audience-gaze on one or the other performer, vs. their instruments, and follow shifts in audience attention between these elements. When this “canjo” tableau ends, the two performers spread out and we see the gaze follow them around the stage. Once one of the performers starts speaking (tableau: “pod rattle incident”), the audience mostly maintain their focus on him, rarely glancing toward the other performer who is gently striking metallic objects. This video highlights that the transformed gaze reliably follows the two musicians around the stage and provides proof-of-concept that our homography approach successfully projects gaze from a large audience in challenging scenes. 

We analyse the 2D distribution of gaze with the help of heatmaps (see Figures \ref{heatmap_ego} and \ref{heatmap_hom}). Figure \ref{heatmap_ego} shows the raw egocentric gaze data recorded from the eye-tracking glasses in the {egoview} coordinate space. This egocentric gaze data is relative to the wearer’s head movements and exhibits low variance in both horizontal and vertical directions with a strong central bias. This effect could be explained by the high coupling between eye and head movements commonly observed in real-world exploration \cite{einhauser2007human, tatler2007central}.

To study the gaze patterns over an evolving scene and compare gaze data between different viewers, we map gaze from each viewer’s frame of reference to the scene’s frame of reference with our Homography module (see Section \ref{Homography}). Figure \ref{heatmap_hom} presents individual gaze heatmaps from all participants and the aggregate gaze activity from all participants for each session overlaid on a reference frame from that session. Projecting the gaze to the static centerview perspective gets rid of the independent positions and head movements of each participant, thereby getting rid of the central bias. Heatmaps of transformed gaze coordinates (Figure \ref{heatmap_hom}A) demonstrate higher gaze dispersion and multiple discrete clusters for each session duration. These clusters coincide with the presentation screen area for the film viewing sessions and the performers’ movement activity for the concert performance sessions (see averaged heatmaps for each session in Figure \ref{heatmap_hom}B). Higher gaze dispersion can be observed during performance viewing than during film viewing, on both days, which could be partially explained by the larger physical space covered on the stage by the performers during the concert, in comparison to the presentation screen size for the film.

Figure \ref{heatmap_avg} highlights different trends in the position of the heatmap centroid (i.e., the peak heatmap intensity location; marked by the intersection of white dashed lines), based on seating position. Investigating the gaze heatmaps for different participant seating clusters, we observe that participants in the right half of the audience spent more time exploring the right side of the stage, reflected by larger (higher gaze dispersion) and more intense (higher gaze duration) clusters in the right half and vice versa (see Figure \ref{heatmap_avg}, panels A and C). This effect is, however, only observed in the performance sessions and not when the participants viewed the pre-recorded documentary film. The heatmaps could further be aggregated to observe similar effects with different participant clusters, such as clustering by participant columns (Figure \ref{heatmap_avg}A) or rows (Figure \ref{heatmap_avg}B). 

\subsubsection{Analysis}

\begin{figure}[htbp]
  \centering
  \includegraphics[width=0.85\linewidth]{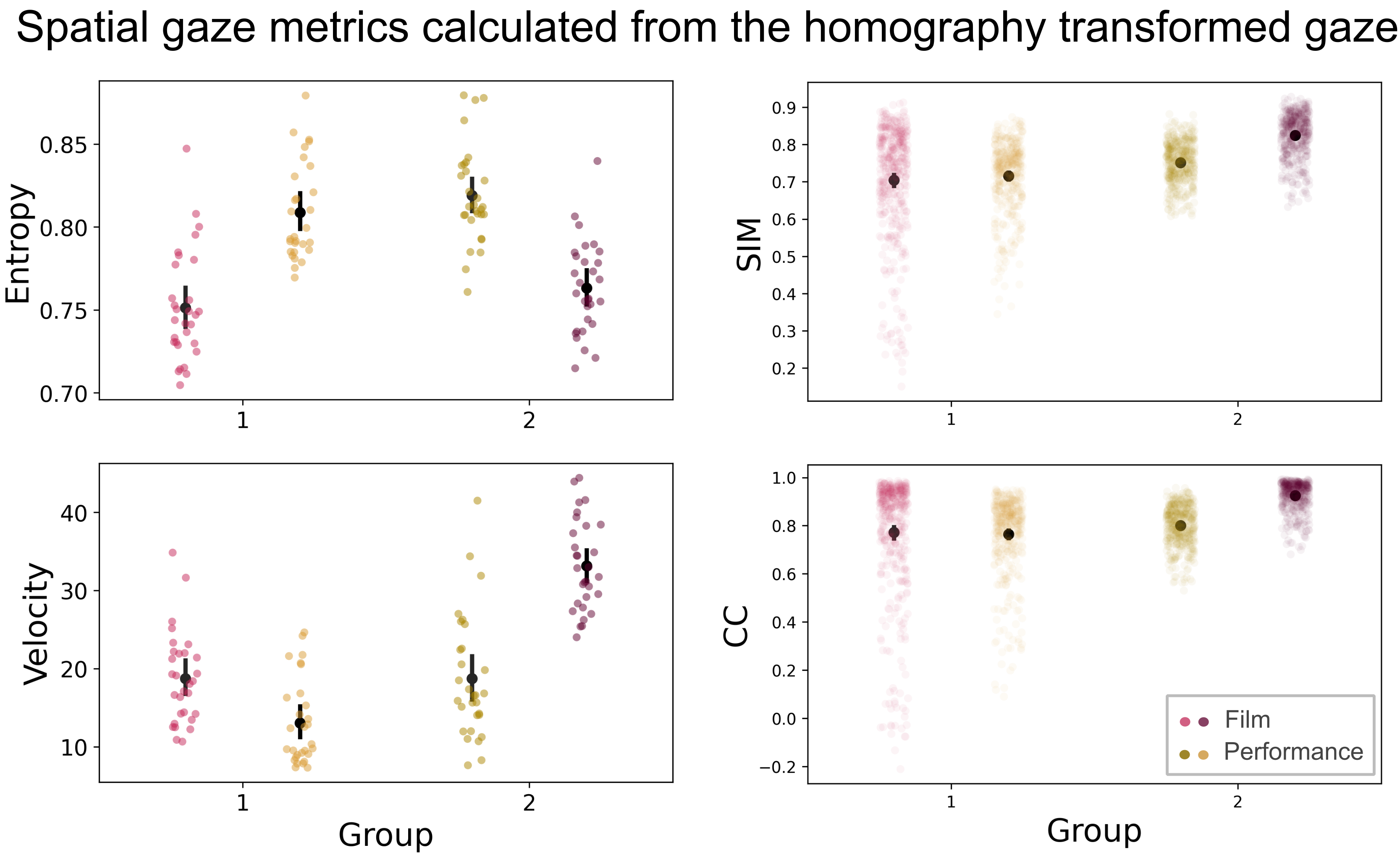}
  \caption{Gaze entropy (\textbf{top left}), average gaze velocity (\textbf{bottom left}), pairwise heatmap similarity (\textbf{top right}) and pairwise heatmap correlation (\textbf{bottom right}) measures computed for each of the two sessions on both days. Coloured dots in the left plots represent individual participant-wise entropy and velocity measures. Coloured dots in the right plots represent SIM and CC for all unique participant-pairs. Black dots in all plots represent means and error bars represent confidence intervals for the respective mean values. Gaze entropy and velocity give a measure of individual gaze variability for each participant and heatmap similarity (SIM) and correlation (CC) provide an estimate of similarities within participants’ gaze exploration for each session.}
  \Description[Spatial gaze measures]{Figure \ref{spatial_measures}. Overall, the gaze entropy is higher and the gaze velocity is lower for performance sessions, as compared to the film sessions. However, further analysis is required to confirm these trends.}
  \label{spatial_measures}
\end{figure}

Further comparisons between individual heatmaps can be made with the help of metrics from our Analysis module (see Section \ref{Analysis}). Pairwise SIM and CC scores were calculated for each participant pair for each of the four event sessions. For raw egocentric gaze, the mean SIM score during film viewing was 0.35 (SD = 0.21) for G1F and 0.38 (SD = 0.19) for G2F, while the mean SIM during the performance was 0.40 (SD = 0.18) for G1P  and 0.43 (SD = 0.17) for G2P. The mean CC was 0.37 (SD = 0.30) for G1F, 0.42 (SD = 0.28) for G2P, 0.43 (SD = 0.26) for G1P and 0.46 (SD = 0.25) for G2P. When using homography-transformed gaze, the scores in all sessions, for both of these metrics, increase drastically (see Figure \ref{spatial_measures}; left column), nearly doubling in most cases. A 2D chart of seat positions (row, column) for each participant (see Figures \ref{heatmap_ego} and \ref{heatmap_hom} for audience seating positions on each day) was used to calculate euclidean distances (DI) between each participant pair. Calculated SIM scores on the homography-transformed gaze were found to be negatively correlated with the euclidean distances during the performance session on both days (i.e., {the closer two participants were to each other, the stronger was the correlation between their gaze patterns} and vice versa). The Pearson correlation coefficient between SIM and DI was -0.10 (\textit{p} = 0.039) for G1P and  -0.17 (\textit{p} < 0.001) for G2P. 

These comparisons highlight the critical role of the homography transformation proposed in our framework for a rich analysis of multi-person eye-tracking datasets. The trends and observations revealed by the transformed gaze data are not available from the raw egocentric gaze; gaze dispersion similarity between participants is much lower in the independent egocentric gaze coordinates and does not highlight clear patterns between participant pairs or groups.

The mean gaze entropy over all participants during film viewing was 0.76 (SD = 0.03) for G1F and 0.77 (SD = 0.03) for G2F. The mean entropy was higher when participants watched the performance, with a value of 0.82 (SD = 0.03) for G1P and 0.83 (SD = 0.03) for G2P. The higher values represent more spatial gaze dispersion when participants watch the performance as compared to the documentary film. 
The average gaze velocity (calculated in pixels/sample) during film viewing was found to be 22.99 (SD = 7.25) for G1P and 36.00 (SD = 5.90) for G2P. During the performance sessions, the average gaze velocity was 16.76 (SD = 6.79) for G1P and 22.11 (SD = 8.03) for G2P. Interestingly, gaze velocity was higher during film viewing on both days which signifies longer saccadic movements when viewing pre-recorded media on a screen, in comparison to a live concert performance.

\begin{figure}[htbp]
  \centering
  \includegraphics[width=0.9\linewidth]{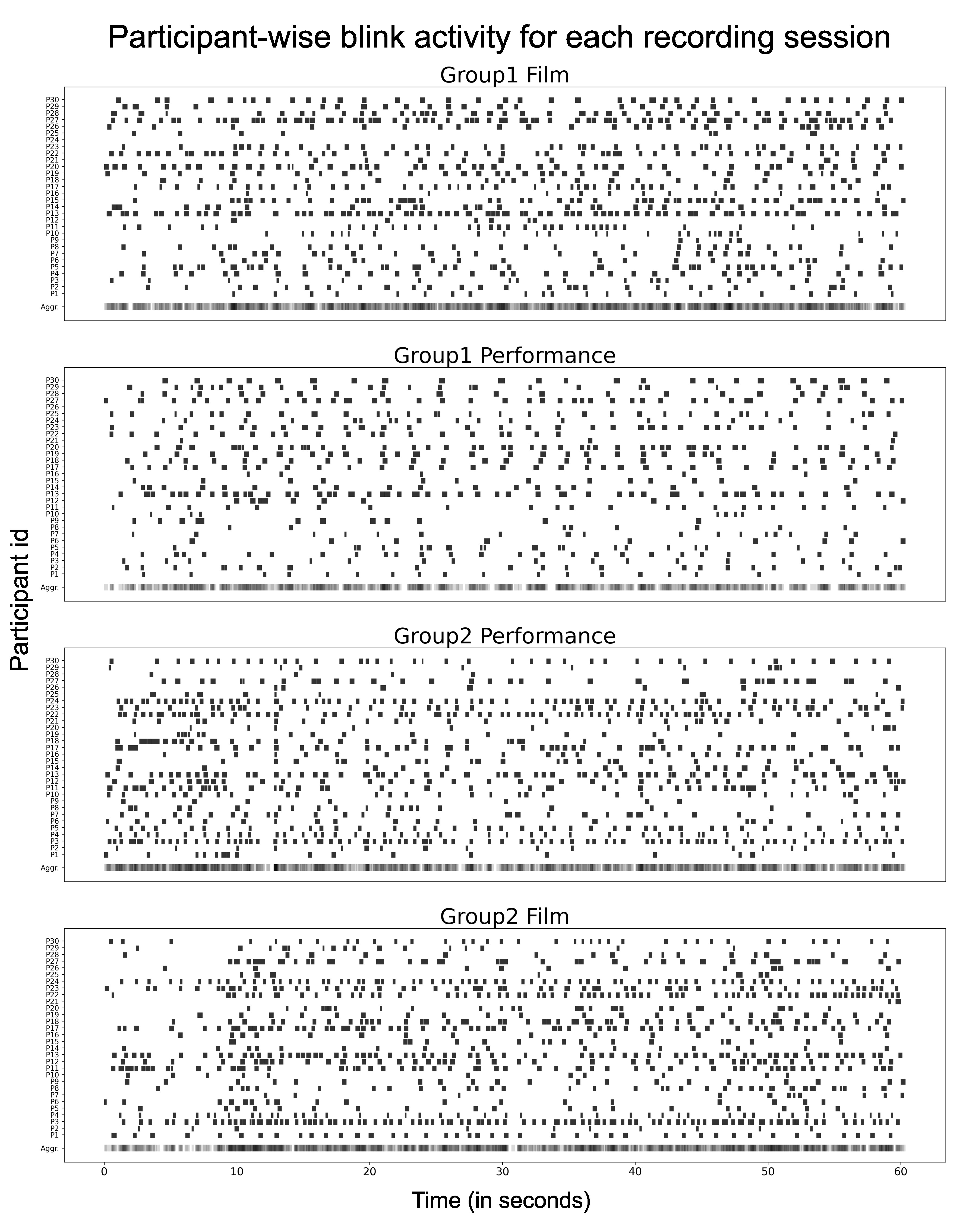}
  \caption{Blinks recorded from individual participants during a one-minute segment of each of the four sessions. Each horizontal bar represents the start and end of a blink with the length representing the blink duration. The bottom row in each of the four plots represents a monochrome heatmap of the overall blink activity over time with darker regions representing more people blinking and lighter regions representing less people blinking at that instant.}
  \Description[Blinks]{Figure \ref{blinks}. The plots demonstrate individual as well as combined blink activity of the audience members over time, which can be directly compared with the audiovisual stimulus presented at the moment.}
  \label{blinks}
\end{figure}

\subsubsection{Exploratory Temporal Analysis}
A major contribution of SocialEyes is the ability to study ocular activity over extended durations, as opposed to controlled short-duration trials in lab experimentation. It allows investigating long-term effects in complex tasks and capturing group dynamics in naturalistic social interactions. Leveraging the precise time synchronisation achieved with our framework, instantaneous ocular activity from multiple people can be investigated. For example, though we can report that the average blink duration for film is 302.93 (SD = 68.55) ms and performance is 306.49 (SD = 69.55) ms, it is much more interesting to be able to analyse blinks in a time resolved manner \cite{lange2023blink}. Figure \ref{blinks} demonstrates blinks from all 30 participants in each of the four recording sessions at discrete time points, over a randomly-clipped 1 minute section. It highlights the variability of individual blink frequency and duration between participants in each session as well as the aggregate blink activity changes with time as each session progresses. Analysing such activity over the entire film or performance (each an hour+ in duration), will offer rich insights into collective information chunking, event segmentation, fatigue, immersion, etc.

\begin{figure}[htbp]
  \centering
  \includegraphics[width=0.9\linewidth]{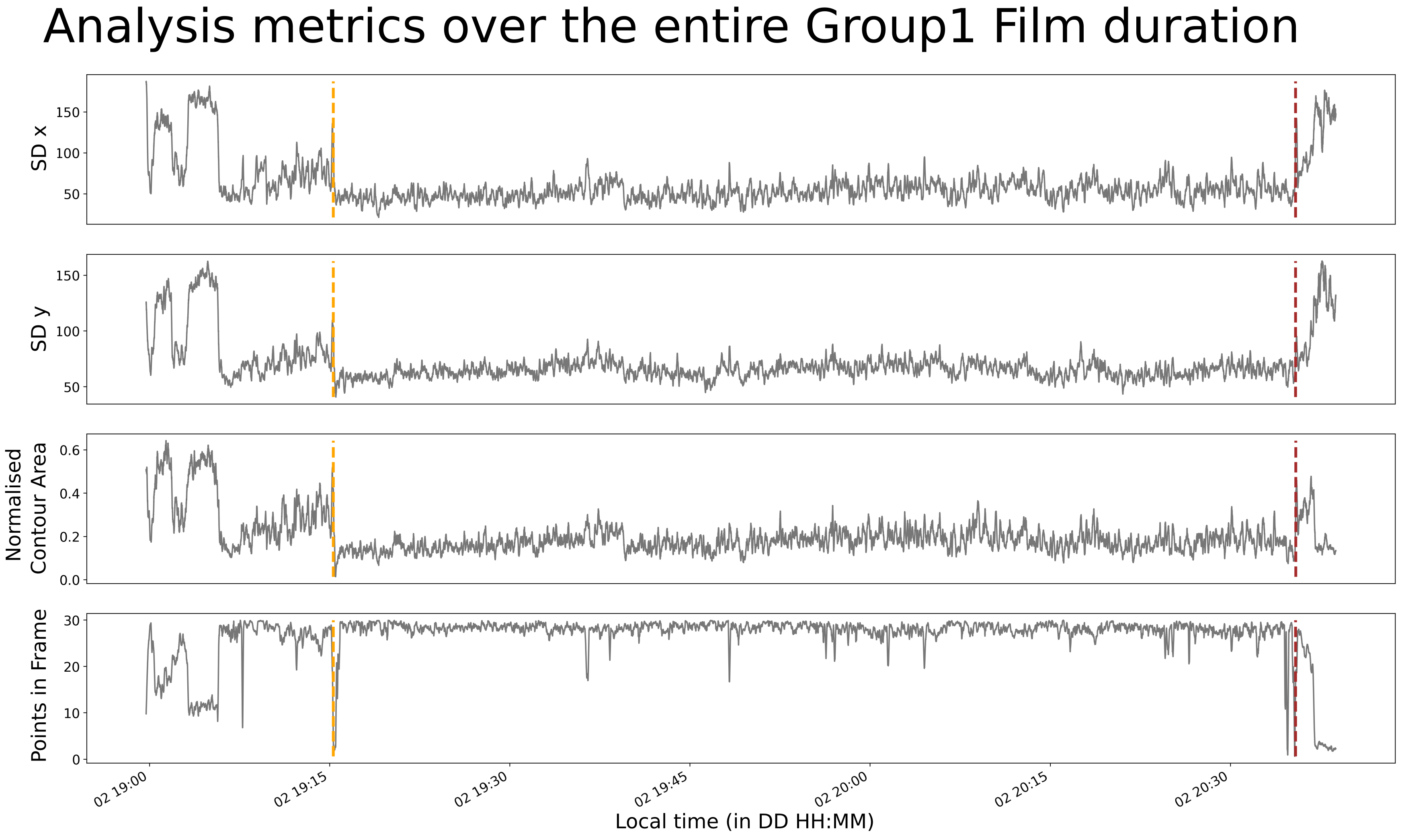}
  \caption{Collective gaze measures over the entire duration of one of the sessions (film viewing on day one). The orange vertical lines represent the start of the film; brown represents the end. The measures are computed by aggregating gaze points from all 30 participants for each frame in the centralview. From top to bottom, the measures are the standard deviation of points in the horizontal direction, standard deviation in the vertical direction, area of a minimum convex hull enclosing all points, and number of points within the centralview frame boundaries. All time series are smoothed with a rolling mean window size of 150 samples to filter high frequency noise.}
  \Description[Temporal metrics]{Figure \ref{temporal_metrics_1}. Fully described in the text.}
  \label{temporal_metrics_1}
\end{figure}

\begin{figure*}[htbp]
  \centering
  \includegraphics[width=0.75\linewidth]{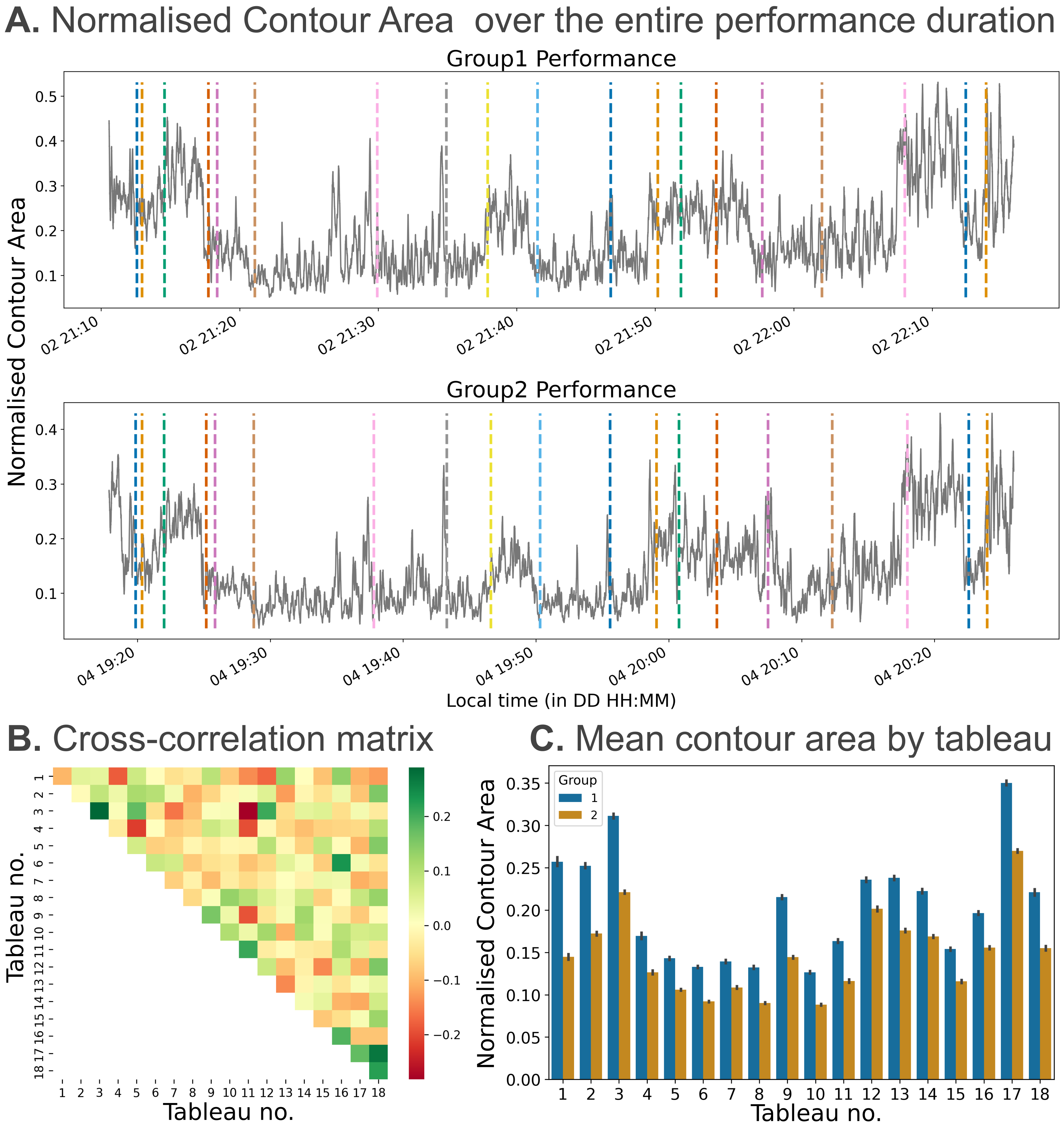}
  \caption{A. Normalised contour area of the minimum convex hull enclosing all gaze points at each time point during the concert performance on both days (G1P: top; G2P: bottom). The performance consisted of 17 different tableaux; the dashed lines represent manually-annotated start times of each tableau in the performance sequentially with additional first and last lines representing start (performers entering stage) and end (end of last tableau) of the performance respectively. The colours correspond to the tableau number. All time series are smoothed with a rolling mean window size of 150 samples to filter high frequency noise for visualisation. B. Cross-Correlation matrix between each pair of tableaux. Green represents higher correlation between the normalised contour area metric during the two tableaux and red represents low correlation. C. Mean contour area of the time series segment for each tableau. The changes in the mean values are similar across the two days (G1P: Blue; G2P: Orange).}
  \Description[Gaze behaviour during concert viewing]{Figure \ref{temporal_metrics_2}. Fully described in the text.}
  \label{temporal_metrics_2}
\end{figure*}

The synchronised gaze data from multiple devices mapped to a common coordinate space (centralview) further facilitates novel spatiotemporal investigations of collective gaze measures over time. In Figure \ref{temporal_metrics_1}, we visualise the measures of standard deviation, minimum contour area, and points in frame, over the entire duration of a film-viewing session (1 hr 20 min film, plus some minutes before/after) from 30 people (G1). These measures–computed by the Analysis module (see Section \ref{Analysis})--reveal collective gaze dynamics over time. Standard deviation of the gaze distribution of all participants at any given time is represented in the horizontal and vertical directions by the "SD x" and "SD y" measures (top and second panels, respectively). The "Normalised contour area" (third panel) represents overall gaze dispersion as the area of a minimum convex hull enclosing gaze from all participants. The number of gaze points falling within the centralview scene at a given timepoint is represented by the "Points in Frame" measure (bottom panel). The time series of all four measures highlight clear event boundaries such as the start and stop of the film presentation (vertical dotted lines). Future analyses will explore which features of the film drive changes in these collective gaze dynamics. 

In another example, visualising the Normalised contour area measure for the entire concert performance (Figure \ref{temporal_metrics_2}) shows similar collective gaze behaviour from audiences during each performance tableau (represented by vertical colored lines) for each participant group. These collective gaze measures and corresponding time series could be further analysed to investigate temporal interactions with the audiovisual stimulus features (outside the scope of the current analysis). We provide these examples here to highlight the rich and vast amount of possibilities from shared gaze metrics in dynamic social settings.

\section{General Discussion}
\label{general-discussion}

Results from our utility test demonstrate the end-to-end implementation of the recording mode of SocialEyes in a real-world application. SocialEyes tackles 1) critical challenges of time synchronisation of multi-sensor data to scale data collection, 2) monitoring and troubleshooting to ensure reliable data collection over long durations, 3) semantic gaze mapping of the synchronised gaze data to analyse collective gaze patterns, and finally, 4) shared gaze visualisation and analysis to achieve novel insights from the collected data. The proposed framework aims to reduce technological and logistical barriers in conducting naturalistic eye-tracking experiments with large social groups, allowing replication/refutation and extension of previous lab studies in social, naturalistic contexts \cite{kulke2021implicit, laidlaw2016camouflaged, foulsham2017fixations}. 

The proposed method conceptualised and tackled key challenges related to multi-person eye-tracking, yet a major external dependency of our framework is the choice of eye-tracking hardware. We acknowledge that the availability/development of required functionality in existing eye trackers and affordability are still substantial barriers. We chose Pupil Labs Neon devices in our utility test over market competitors because of their remote API control, minimal calibration requirements, and the availability of open-source gaze-processing components. However, in our initial tests we encountered an array of operational issues on the proprietary vendor application that severely hindered reliable recording from the devices. Solving these issues required hours of testing, troubleshooting or downscaling. For example, the Neon devices are also capable of recording device orientation and movement at 110 Hz using the onboard 9-DoF Inertial Measurement Unit (IMU), as well as stereo audio from on-board microphones; we had to omit these data streams to allow reliable recording of the gaze and video data streams, which were our priority. IMU sensor failures would crash the mobile application, even though the eye data were intact, and audio recording could not be reliably started remotely through API calls, as it requires the Android device to be unlocked. {While these issues are actively being resolved with the manufacturer's assistance, they point out the very incipient stage of multi-person eye-tracking hardware/software, which are, traditionally, only tested for single person use cases.} Moving forward, the technology would benefit from rigorous testing at the scale we propose in this paper. Further, the availability of open-source implementations would allow higher customizability for application specific requirements and thereby contribute to its adoption in novel settings.

\subsection{{Limitations and Future Directions}}

\subsubsection{{Edge-cases in gaze projection}}
\label{edge-cases}

{
Future objectives for our framework would be to further investigate and optimise gaze mapping accuracy in different environments. In our previous pilot study \cite{Saxena_SVS_poster}, we identified a few edge case scenarios that pose a challenge to the homography transformations. In summary, the SuperGlue + RANSAC approach demonstrated robustness to head movements, illumination changes, and scene occlusions, as observed in our current utility study as well. However, close distances between the observer and scene as well as duplicate objects in a scene can result in false predictions. Systematically identifying these edge failure cases in challenging environments and addressing them—either through software and algorithmic updates or by establishing well-defined best practices for experimental setups—could significantly improve the robustness and reliability of gaze projections. To automate the collection of ground-truth gaze locations in these investigations, fiducial markers can be used by guiding participants to fixate on specific targets–as demonstrated in Appendix \ref{Appendix}.}

\subsubsection{{Alternatives to planar homography}}
\label{homography_alternatives}

{
The assumption of a substantial distance between the observer(s) and the scene is a theoretical limitation of the current homography-based gaze projection as it projects a plane at sufficiently large (infinite) distance from one view to another. This assumption allows for more reliable and calibration-free projections, particularly when dealing with moving cameras, such as the head-mounted egoview, and dynamic (non-static) scenes, such as a music performance. We show the validity of this assumption for our current setup in Appendix \ref{Appendix}. Future optimisations can also consider multi-view homography where the scene is captured by multiple centralview cameras and each egoview camera is related with the centralview corresponding to its position. This strategy would enable more precise cross-view analyses and spatial consistency in wide-view environments. To eliminate the 2D plane assumption, two alternative approaches are possible. Given the intrinsic parameters of the two cameras (egoview and centralview) and a common 3D coordinate system, a 2D point from the egoview can be projected to the 3D world coordinate system and then to the 2D centralview. This approach relies on pre-existing information about the locations of objects in the 3D world space, which is highly unfeasible, particularly in dynamic scenes. The second approach involves calibrating extrinsic positions of each egoview-centralview camera pair and projecting the 2D point from one camera's view to another using epipolar geometry. This triangulation approach provides a general (ideal) solution, particularly for scenes with varying depths or close viewing distances where the homography planar assumption is not valid. However, it relies on accurate calibration parameters between the two cameras which is challenging with non-stationary cameras such as the head-mounted egoview cameras. As a result, the application of this approach relies on accurate and frequent re-calibration of the egoview cameras and prominent (unobstructed, well-lighted, strategically placed) visual markers that can be detected in the dynamic scene. These requirements might often interfere with the experimental setup, since the control of lighting and marker placement is not always possible. Additionally, markers create a visual distraction for viewers that would interfere with scientific investigations of visual attention. As an alternative to visual markers, localisation of the egoview camera (and therefore the eye-tracker) is also possible with invisible infrared markers \cite{infraredTags}, feature-based tracking via algorithms such as visualSLAM \cite{visualSLAM} and optical flow \cite{opticalflow}, marker-less methods \cite{stageAR}, or using the on-device 9DOF Inertial Measurement Unit (IMU) sensors. However, each method has limitations, such as marker detection robustness over distance, the need for a pre-registered 3D world model in model-based approaches, and the reliability of the IMU sensor with the existing eye-tracker (see Section \ref{general-discussion}). Overall, triangulation based methods might offer benefits, such as depth compensation and runtime efficiency, warranting future investigations to evaluate their trade-offs in performance, accuracy, and ecological validity compared to the current homography approach.}

\subsubsection{{Optimisations for Streaming mode of operation}}

{
The current utility study evaluated the recording (offline) mode of SocialEyes (defined in Section \ref{Modes of operation}) where accuracy and resolution of recorded data are prioritised over run-time efficiency. While certain components of the recording mode, like real-time device monitoring, are directly transferable to the streaming (online) mode, future optimisations for real-time gaze projection, data acquisition/transmission, and hardware-accelerated video decoding for multi-camera streaming need to be tested at a scale similar to our current test. Below, we share insights and recommendations on these optimisations based on the current testing experience.}

{
Real-time gaze projection could be realised with the existing SuperGlue + RANSAC method or the alternative approaches discussed in Section \ref{homography_alternatives}, tuning for the desirable accuracy-performance trade-off. For instance, marker-based plane detection or feature tracking using methods like Kanade–Lucas–Tomasi \cite{lktTracker} can offer better run-time efficiency with slightly less accuracy or reliability of projective transformation. Similarly, multi-view homography and stereo calibration based 3D triangulation can offer better accuracy and robustness in diverse environments but might be more computationally intensive.}

{
Data storage and transmission could be optimised in recording mode using the IMU data by only retaining the frames corresponding to substantial rotational or translational changes. These motion changes can also be detected from the egoview camera signal by calculating pixelwise change or optical flow, which reduces the requirement of an additional IMU data stream. However, detecting motion from pixels as opposed to IMU sensors would also be more sensitive to video artefacts such as lighting changes. In contrast, for the recording mode used in the current study, we ensured data reliability and reduced network utilisation by storing data locally on each eye-tracking device. Nonetheless, these strategies of motion detection might still be applied to optimise offline computation time by detecting robust features only on key frames with high inter-frame motion and tracking them on the remaining frames with low motion.}

{
Streaming video data concurrently from multiple egoview cameras is typically challenging with software decoding due to increased load on CPU. For instance, in our current eye-tracking hardware, egoviews are sent as Full HD H.264 streams over the network. Our initial tests on a server with an Intel Core i9-13900KF CPU and a NVIDIA GeForce RTX 4090 GPU resulted in latency issues and dropped key-frames when attempting more than 5 concurrent video streams decoded on CPU (default manufacturer implementation). However, hardware-accelerated decoding with NVDEC (custom implementation with FFMPEG) was able to reliably stream all 30 egoviews concurrently with 10-30\% GPU utilisation. Future testing for longer durations and optimizations, such as reducing stream bitrates—which are currently constrained by the proprietary nature of the eye-tracking application—and refining the decoding methods, will be crucial to ensure concurrent stream decoding.}


{\subsubsection{Scope enhancement}
Finally, the SocialEyes framework can be extended to support additional social environments and hardware vendors.} The current paper demonstrates gaze mapping in scenes where multiple people share a scene view; however, the application could be scaled up to social settings where individuals share different views of the same scene using Multi-view Multi-object detection and tracking \cite{taj2010multi}. {Creative applications, such as gaze-contingent interactions and} sonification \cite{hornof2014creative2}, provide an exciting avenue for exploration as well. Future development and applications can also incorporate multiple sensor types, such as physiological or IOT sensors, to record and analyse synchronous multimodal measures. We look forward to applying this framework to study attention and subjective experiences in immersive, social contexts, and open-source the code to motivate collaboration and application of our framework in novel contexts.

\section{Conclusion}
This paper presents SocialEyes--an accurate and scalable framework to conduct multi-person eye-tracking in everyday social situations where people have a shared gaze goal. In the current study, we used a live music concert and a film screening, but the described approach is generalizable. Building on previous studies that have identified eye measures to reflect visual and auditory attention, our proposed framework allows studying such multimodal correspondences outside of a controlled lab environment, and with multiple participants simultaneously. Such multi-person eye-tracking in dynamic social settings promises to provide novel insights into group behaviour, new approaches to cooperative/collaborative work in large-group settings, and new means to create gaze-contingent immersive environments and social interactions.  

\section{Code Availability}

We have made the source code for our SocialEyes implementation, as well as the scripts for generating results and figures from the Utility Test (Section \ref{Utility Test}), available to promote open research and reproducibility at the following link: \url{https://github.com/beatlab-mcmaster/SocialEyes}.

\begin{acks}

We would like to thank percussionist-composers Allen Otte and John Lane for creating and performing \emph{The Innocents} concert, along with director Wojciech Lorenc for his work on the accompanying film (see Section \ref{Stimuli}). Their artistry and vision provide a compelling foundation for this research and future studies on the subject.

This work was supported by the \grantsponsor{}{Natural Sciences and Engineering Research Council of Canada}{} Discovery Grant [\grantnum[]{}{RGPIN-2023-05050}], \grantsponsor{}{the Canadian Foundation for Innovation}{} John R. Evans Leaders Fund grant [Project \grantnum[]{}{\#43884}], and the \grantsponsor{}{Ontario Research Fund}{} for Small Infrastructure [\grantnum[]{}{\#43884}] held by LF. We would also like to thank the German Academic Exchange Service (DAAD) for providing a scholarship to AN, which supported his participation in this research at McMaster.

\end{acks}

\bibliographystyle{ACM-Reference-Format}
\bibliography{main}

\appendix 
\section{{Appendix: Evaluation of homography projection at varying scene depths}}
\label{Appendix}
{
In the current study, gaze projection is achieved using homography, which projects the visual scene on stage as a 2D plane from the egoview to the centralview perspective. This planar mapping assumes that the scene can be approximated as a plane at a sufficiently large (effectively infinite) distance from the camera. To assess the validity of this plane-at-infinity assumption, we conducted a post-hoc validation procedure. Note that this procedure and associated results do not directly relate to the performance and design of the SocialEyes framework presented in this paper. The aim here is to quantify the impact of varying scene depths on the accuracy of the current homography-based projections, providing additional insights for interested readers.}

\begin{figure}[htbp]
  \centering
  \includegraphics[width=\linewidth]{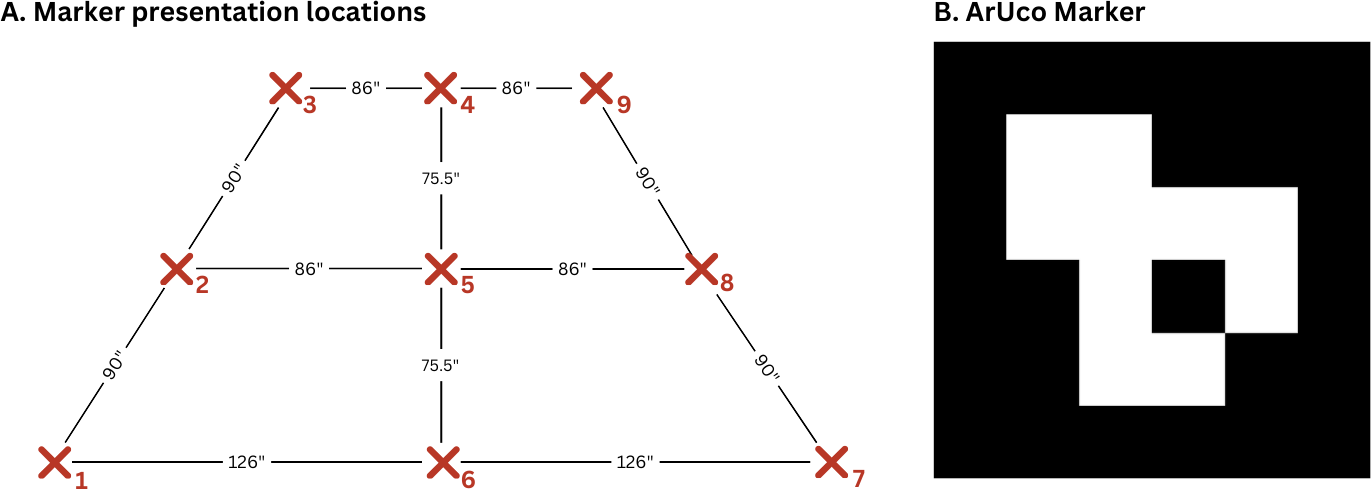}
  \caption{A. The visual marker was placed at 9 different locations spread accross the stage. Red X's in the image mark these positions. Distances between the positions are marked in inches. The top of the image represents the back of the stage and the bottom represents the front. Subscripted numbers 1-9 represent the sequence of presentation. B. An image of the visual marker used in the study.}
  \Description[Visual marker placement and design]{Figure \ref{appendix_apparatus}. The placement positions in A form an isosceles trapezium, with the larger base facing the audience. The marker was positioned at each vertex, the midpoint of each edge, and the center of the trapezium. The presented marker in B is ID 9 from a 4×4 ArUco dictionary.}
  \label{appendix_apparatus}
\end{figure}

\subsection{{Method}}

\subsubsection{{Apparatus}}
{
The validation procedure was conducted with the existing Pupil Labs Neon eye-trackers in the same physical space as the Utility Test (Section \ref{Utility Test} of the main paper). Centralview was captured with an ArduCam EK035 camera, featuring a global shutter and frame-wise hardware timestamping. An ArUco marker--25.5 inches wide and 21 inches tall--was mounted on a sturdy stand and displayed to the audience at 9 different positions on the stage. Neon devices were wirelessly connected to the local network and the current SocialEyes implementation was used for data collection and analysis.}

\subsubsection{{Procedure}}
{
19 participants were recruited during a public event at LIVELab, McMaster University. Participants seated themselves randomly among the audience seats and were equipped with the eye-trackers before the event started. The seating rows began at a distance of 134 inches from the stage front, with 50 inches of gap between each consecutive row. The validation procedure was conducted at the end of the hour-long event and was completed in roughly 3 minutes. During the procedure, an experimenter serially placed the ArUco marker (Figure \ref{appendix_apparatus}B) at each of the positions--$X_{1
}$
 to $X_{9}$--presented in Figure \ref{appendix_apparatus}A. Participants were instructed to look at the ArUco target while we counted down from five. Doing so ensured that the target was visible in their egoviews.}

\subsubsection{{Data Analysis}}
{
100 consecutive frames from the centralview video were selected for each of the 9 presentation locations. ArUco markers were detected in time-synchronised egoviews of each participant and transformed to centralview using SocialEyes' Homography module (Section \ref{Homography}). Projection errors were calculated as the Euclidean distances between the transformed and the detected marker centers in the centralview (640x480 resolution). Data from five devices were excluded due to more than 40\% missing projections, either due to failure in marker detection or due to an inadequate number of features detected for homography.}

\subsection{{Results}} 
{
Errors for each presentation location and depth are presented in Tables \ref{tab:appendix_pt_results} and \ref{tab:appendix_depth_results} respectively.  Aggregated mean projection error reduced with an increase in depth, however, pairwise comparisons using a Wilcoxon signed-rank test showed no significant differences between depth levels 1 and 2 (W = 86.00, p = 1.00), 1 and 3 (W = 39.00, p = 0.07), or 2 and 3 (W = 44.00, p = 0.13), where depth level 1 is closest to the audience and 3 is the farthest.}

\begin{table}[htbp] 
    \centering
        \centering
        \begin{tabular}{|c|c|c|c|}
            \hline
            \textbf{{Position}} & \textbf{{Frames}} & \textbf{{Mean}} & \textbf{{Std}} \\ \hline
            1 & 882 & 47.83 & 98.94 \\
            2 & 1090 & 49.35 & 104.60 \\
            3 & 1106 & 32.93 & 39.82 \\
            4 & 1141 & 29.62 & 41.65 \\
            5 & 1248 & 41.87 & 74.26 \\
            6 & 1303 & 55.22 & 45.03 \\
            7 & 962 & 39.44 & 44.43 \\
            8 & 1184 & 30.29 & 33.84 \\
            9 & 969 & 35.01 & 32.84 \\
            \hline
        \end{tabular}
        \caption{{Errors calculated for each presentation location. The position indices correspond to positions in Figure \ref{appendix_apparatus}.}}
        \label{tab:appendix_pt_results}
    \end{table}
\begin{table}[htbp]
        \centering
        \begin{tabular}{|c|c|c|c|}
            \hline
            \textbf{{Depth}} & \textbf{{Frames}} & \textbf{{Mean}} & \textbf{{Std}} \\ 
            \hline
            1 & 3147 & 48.32 & 65.02 \\
            2 & 3522 & 40.29 & 76.04 \\
            3 & 3216 & 32.38 & 38.60 \\ 
            \hline
        \end{tabular}
        \caption{{Errors aggregated for each of the 3 depth levels. Depth level 1 was closest to the audience (front of stage) and 3 was the farthest (back of stage).}}
        \label{tab:appendix_depth_results}
\end{table}

\subsection{{Discussion}} 

This validation procedure evaluated the projection error using our proposed homography approach, at varying scene depths. A fiducial marker was recorded from multiple egoviews spread across the audience seating space and the projected marker-centre from each egoview was compared to the ground-truth marker-centre in centralview. Projection errors across three different scene depths (each comprising of varying viewing angles) revealed no significant differences suggesting that the homography planar assumption is valid for the current setup. 

A limitation in the current procedure is the poor image quality of the centralview camera due to the unavailability of appropriate Image Signal Processing (ISP) techniques for the utilised sensor. We used the ArduCam camera in this procedure for its benefits of global shutter and hardware timestamping that resulted in precise time synchronisation, however, the lack of ISP configuration for the sensor resulted in added noise to the homography projections. Note that this centralview camera was different from the one used in the Utility Study reported in the main paper, which did not have any ISP issue.

In summary, despite the added sensor noise in centralview data, the homography planar assumption remains valid. The projection errors did not show significant variation across different scene depths. For future applications, it is important to ensure sufficient distance between the scene and observers for the assumption to hold, as projection errors may increase significantly at shorter distances.

\end{document}